\documentclass[epj,twocolumn]{svjour}
\usepackage{cite,mcite}
\usepackage{graphicx}
\usepackage{subfigure}
\usepackage{atlasphysics}
\usepackage{amssymb}
\usepackage{hyperref}
\usepackage{helvet}
\usepackage{lineno}

\def\thedate{October 30, 2011}

\def\Pythia{{\sc Pythia}}

\def\Herwig{{\sc Herwig}}
\def\Sherpa{{\sc Sherpa}}
\def\Perugiatwenty{{\sc Perugia2010}}
\def\Herwigpp{{Herwig\raise0.2ex\hbox{\scriptsize\kern-0.5pt+\kern-0.5pt+}}}

\def\antikt{anti-$k_t$}

\def\GeV{{\rm GeV}}
\def\MeV{{\rm MeV}}
\def\TeV{{\rm TeV}}

\def\pbi{{\rm pb}^{-1}}

\def\ptTrack{\ensuremath{p_{\mathrm{T\,track}}}}

\def\etaJet{\ensuremath{\eta_{\mathrm jet}}}

\def\ptJet{\ensuremath{p_{\mathrm{T\,jet}}}}
\def\ntrk{\ensuremath{N_{ch}}}
\def\njet{\ensuremath{N_{\mathrm jet}}}
\def\ptrel{\ensuremath{p_{\mathrm{T}}^{rel}}}

\def\rhoch{\ensuremath{\rho_{ch}(r)}}

\def\simge{
    \mathrel{\rlap{\raise 0.511ex
        \hbox{$>$}}{\lower 0.511ex \hbox{$\sim$}}}}
\def\simle{
    \mathrel{\rlap{\raise 0.511ex
        \hbox{$<$}}{\lower 0.511ex \hbox{$\sim$}}}}
\def\abs#1{\left| #1\right|}
\begin{document}
\title{\vspace{-3.0cm}
\flushleft{\normalsize\normalfont{CERN-PH-EP-2011-148}} 
\flushright{\vspace{-0.86cm}\normalsize\normalfont{Submitted to Eur. Phys. J. C}} \\[3.0cm] 
\flushleft{Measurement of the jet fragmentation function 
and transverse profile
in proton-proton collisions at a
  center-of-mass energy of 7~TeV with the ATLAS detector}}
\titlerunning{Jet fragmentation function and transverse profile}
\author{The ATLAS Collaboration}
\date{\thedate}
\institute{}
\abstract{
The jet fragmentation function and transverse profile for
jets with $25\ \GeV < \ptJet\ < 500\ \GeV$ and $|\etaJet|<1.2$ produced in 
proton-proton collisions 
with a center-of-mass energy of 7~\TeV\ are presented.
The measurement is performed using data with 
an integrated luminosity of $36\,\pbi $.
Jets are reconstructed and their momentum measured using calorimetric
information.  The momenta of the charged particle constituents
are measured using the tracking system.  
The distributions corrected for detector effects are
compared with various Monte Carlo event generators and generator tunes.
Several of these choices show
good agreement with the measured fragmentation function.
None of these choices reproduce both the transverse profile and
fragmentation function  over the full kinematic range of the measurement.
}
\maketitle
\section{Introduction and Overview}
\label{Sec:intro}

This paper presents measurements of jet properties  in 
proton-proton ($pp$)
collisions at a center of mass energy of $7 \ \TeV$ at the CERN
LHC using the ATLAS detector. Jets are identified and their
momenta measured using the calorimeters. Charged particles
measured by the tracking system are then associated with these
jets using a geometric definition. The structure of the jets is 
studied using these associated particles.

Jets produced at large transverse momentum in proton-proton collisions
arise from the scattering of proton constituents leading to outgoing partons 
(quarks and gluons)
with large transverse momenta.
These manifest themselves as jets of hadrons
via a ``fragmentation process''. While the scattering of the proton 
constituents
is well described by perturbative QCD and leads, at lowest order,
to final states of $gg$, $gq$,  and $qq$,  the fragmentation process is 
more complex.
First, fragmentation must connect the outgoing partons with
the rest of the event as the jet consists of colourless hadrons while the
initiating parton carries colour. Second, the process involves the production
of hadrons and takes place at an energy scale where the QCD coupling
constant is large and perturbation theory cannot be used.
Fragmentation is therefore described using a QCD-motivated model with
parameters that must be determined from experiment.
The fragmentation function $D^h_i(z,Q)$ is defined  as the 
probability that a hadron of type $h$ carries longitudinal
momentum fraction $z$ of the momentum $\vec{p}_i$ of a parton of type $i$
\begin{equation}
   z \equiv \frac{\vec{p}_i\cdot\vec{p_{h}}}{\abs{\vec{p}_i}^2}.
\label{eq:1}
\end{equation}
$D(z,Q)$ depends on $z$ and on the scale $Q$ of the hard scattering 
process which produced the parton.
While the value of  $D^h_i(z,Q)$ cannot be calculated in perturbative
QCD, the variation with $Q$ can be predicted provided $Q$ is sufficiently 
large\cite{Gribov:1972rt,Altarelli:1977zs,Dokshitzer:1977sg,Georgi:1977mg,Owens:1978qz,Uematsu:1978yw}.

In this paper  a quantity related to $D^h_i(z,Q)$ is 
measured. After jets have been
reconstructed, the data are binned for fixed ranges of jet transverse 
momenta (\ptJet),
each bin containing \njet\ jets; $z$  is then
determined for each charged particle associated with the jet 
\begin{equation}
z = \frac{\vec{p}_{\mathrm jet}\cdot\vec{p}_{ch}}{\abs{\vec{p}_{\mathrm jet}}^2},
\end{equation}
where $\vec{p}_{\mathrm jet}$ is the momentum of the reconstructed jet and
$\vec{p}_{ch}$ the momentum of the charged particle.
The following quantity is measured
\begin{equation}
F(z,\ptJet)  \equiv \frac{1}{\njet} \frac{d\ntrk}{dz},
\label{eq:2}
\end{equation}
where \ntrk\ is the number of charged particles in the jet.
$F(z,\ptJet)$ is a sum over  $D^h_i(z,Q)$ weighted 
by the rate at which  each parton species ($i$) is produced from the hard scattering process.
As particle identification is not used, $h$ is summed over all charged hadrons.
The hard scattering scale $Q$ is of the same order of magnitude
as \ptJet.
At small \ptJet, gluon jets dominate due to the larger gluon 
parton densities in the proton and larger scattering rates
for $gg\to gg$. In the pseudorapidity range used for jets in this 
analysis ($\abs{\etaJet} < 1.2$)\footnote{ATLAS uses a 
right-handed coordinate system with its origin
at the nominal interaction point (IP) in the centre of the 
detector and the $Z$-axis coinciding with the axis of the beam
pipe. The $X$-axis points from the IP to the centre of the LHC
ring, and the $Y$-axis points upward. Cylindrical coordinates
($r$,$\phi$) are used in the transverse plane, $\phi$ being the azimuthal
angle around the beam pipe. The pseudorapidity is defined in
terms of the polar angle $\theta$ as $\eta=-\ln \tan(\theta/2)$.
The rapidity $y$ for a track or jet is defined by 
$y=0.5\ln \left [ (E+p_Z)/(E-p_Z) \right ] $ 
where $E$ denotes the energy and $p_Z$ is
the momentum along the beam direction.   For tracks, the energy is
calculated assuming the particle is a pion.} 
the fraction of jets originating from
a hard scattering that produces a gluon falls from 80\%\ for  
$\ptJet \sim 25 \ \GeV$ to 50\%\ for  $\ptJet\sim 300 \ \GeV$ according
to the \Pythia~\cite{Sjostrand:2006za} event generator.

The jets measured experimentally also contain particles produced from
the hadronization of the beam remnants (the ``underlying event'').
It should be emphasized that because colour fields connect all the strongly
interacting partons in the $pp$ event, no unambiguous assignment of 
particles
to the hard scattering parton or underlying event is possible.
The integral of
$F(z,\ptJet)$ with respect to  $z$ corresponds to the multiplicity of
charged particles within the jet. A clear summary 
of fragmentation phenomenology
is provided in~\cite{Nakamura:2010zzi}
(Section 17)  whose 
notation
is followed here. 

The derivation of  $D^h_i(z,Q)$ from $F(z,\ptJet)$ is 
beyond the scope
of this paper, but comparisons of  $F(z,\ptJet)$ with the
predictions of several Monte Carlo (MC) generators will be made. Different
features of the  Monte Carlo  models are probed by these studies.
At low values of \ptJet, the comparisons are most sensitive to the
non-perturbative models of fragmentation, the connection of the partons
to the remainder of the event and to the accretion  of
particles from the underlying event into the jet.
As \ptJet\ rises, the impact of these effects is diluted and,
if all the  Monte Carlo models implemented perturbative QCD in the
same way,  $F(z,\ptJet)$ would become similar.
In particular the increase
of the total particle multiplicity with the hard scattering energy, here
 \ptJet,   is predicted by
perturbative QCD\cite{Mueller:1981ex}.

Two other related quantities that describe the transverse shape of the 
jets are also studied here.
The variable \ptrel\ is the momentum of charged 
particles in a jet transverse to that jet's axis:

\begin{equation} \ptrel  = \frac{|\vec{p}_{ch} \times 
\vec{p}_{\mathrm jet}|}{|\vec{p}_{\mathrm jet}|}. \end{equation}
\noindent The following distribution is measured
\begin{equation}
f(\ptrel,\ptJet) \equiv \frac{1}{N_{\mathrm jet}} \frac{d\ntrk }{d\ptrel}.
\label{eq:3}
\end{equation}

\noindent Finally, the density of charged particles in $y$-$\phi$ space,\\ 
$\rho_{ch}(r,\ptJet)$, is
measured as a function of the angular distance $r$ of charged particles 
from the axis of the jet that contains them, where $r$ is given by:

\begin{equation} \label{eqn:rdef} r = \Delta R(ch,jet) = 
\sqrt{(\phi_{ch} - \phi_{\mathrm jet})^2+(y_{ch} - y_{\mathrm jet})^2} 
\end{equation}

\noindent Thus  $\rho_{ch}(r,\ptJet)$ is given by:

\begin{equation} \rho_{ch}(r,\ptJet) \equiv \frac{1}{\njet} 
\frac{dN_{ch}}{2\pi rdr}. 
\label{eq:5}
\end{equation}

As in the case of the longitudinal variables, a comparison of these 
transverse quantities with Monte Carlo
generators is sensitive to many of their features. The non-perturbative 
hadronization processes produce particles  that 
have limited transverse momentum with respect to the parton direction.
The mean value of this transverse momentum is of order
a few hundred \MeV, the scale where the QCD coupling constant becomes 
non-perturbative.
At low \ptJet\ this effect dominates.
If there were no other contributions, 
\ptrel\ would remain constant 
with increasing \ptJet.  Therefore  more of the energy would be 
concentrated in the core of the jet  as  \ptJet\ increases 
and the jets would become narrower.
However, as \ptJet\ increases
contributions from processes controlled by perturbative QCD radiation 
become more 
important, contributing to
jet broadening and causing the
mean value of \ptrel\ to rise slowly (approximately  logarithmically).

The phenomena described above are incorporated in all the Monte Carlo 
generators used to describe jet production in $pp$ collisions, although
there are significant  differences in how these effects are implemented.
For example, \Pythia\ 
describes non-perturbative hadronization using a 
string model while \Herwig~\cite{Corcella:2000bw}  uses a cluster model.  
In \Pythia, coherent
colour effects are described partly by string fragmentation.
These effects are also produced  in \Herwig\ and  \Pythia\
from gluon radiation.  Treatments of the 
proton remnants are also described using different phenomenological
approaches.  For both generators, the implementations require that
a number of input parameters be tuned to the data.  The results presented
in this paper will test whether these Monte Carlo models and their current
input parameters adequately describe jets produced at the LHC.
As the results are presented in bins of  \ptJet, the explicit dependence on
\ptJet  in the variables defined in 
Equations~\ref{eq:2}, \ref{eq:3} and~\ref{eq:5} is
often suppressed in the following.

 The measurement is performed using 
data with an integrated luminosity of $ 36\ \pbi $ 
recorded in 2010 with the ATLAS
detector at the LHC at a center-of-mass energy of 7~$\TeV$.  
The measurement covers a
kinematic range of $25\ \GeV < \ptJet < 500\ \GeV$ and $|\etaJet| < 1.2$. 
Events are triggered using a minimum bias trigger and a
combination of calorimeter jet triggers. A complementary ATLAS 
analysis\cite{Aad:2011gn} studying the jet fragmentation 
function and transverse profile
of jets reconstructed from charged particle
tracks using a total integrated luminosity of ~800 $\mu {\rm b}^{-1}$ has 
been completed. 
It  explores the properties of jets at lower transverse momentum 
than those typically studied in this paper.

Previous measurement of jet fragmentation functions have been made in $e^+e^-$ collisions
  \cite{Akers:1995wt,Abreu:1996na,Buskulic:1995aw,Mattig:1988sv}, in $p\overline{p}$  collisions 
\cite{Arnison:1983gv,Abe:1990ii} and in $ep$ collisions~\cite{Breitweg:1997ra,Aaron:2007ds}.  

This paper is organized as follows.
The ATLAS detector is described briefly in Section~\ref{sec:detector}.
The Monte Carlo generator samples are discussed in Section~\ref{sec:samples}.
The event and object selections are described in Section~\ref{sec:selection}. 
Section \ref{sec:analysis} contains a description of the
analysis. In Section~\ref{sec:systematics} the treatment of systematic 
uncertainties is presented. Results and conclusions are shown in Sections~\ref{sec:results} and~\ref{sec:conclusions}.
\section{The ATLAS Detector}
\label{sec:detector}
The ATLAS detector is described in detail in 
Ref.~\cite{ATLAS}.  The subsystems
relevant for this analysis are the inner detector (ID), 
the calorimeter and the trigger.
The ID is used to measure the momentum of charged particles.
It consists of three subsystems: a pixel detector, a silicon strip tracker
(SCT) and a transition radiation straw tube tracker (TRT).  These
detectors are located
inside a solenoid  that provides a 2~T axial field.  
The ID has full coverage in the azimuthal angle 
$\phi$ and over the pseudorapidity range
$0<|\eta^{\mathrm track}|<2.5$.

The electromagnetic calorimeters use liquid argon as the
active detector medium.  They consist of accordion-shaped 
electrodes and lead absorbers
and  cover the pseudorapidity range $|\eta|<3.2$.
The technology used for the hadronic calorimeters varies with $\eta$.
In the barrel region ($|\eta|<1.7$) the detector is made of 
scintillating tiles with steel radiator.  In the endcap region 
($1.5<|\eta|<3.2$) the detector uses liquid argon and copper.
A forward calorimeter consisting of liquid argon and tungsten/copper absorbers 
serves as both electromagnetic and hadronic calorimeter at large 
pseudorapidity and extends the coverage to  $|\eta| < 4.9$.

The calorimeters are calibrated at the electromagnetic scale 
which correctly reconstructs the energy
deposited by electrons and photons.
The calorimeters are not compensating and  the
response of hadrons is lower than that of electrons ($e/h>1$). 
Some fraction of the 
hadronic energy can also be deposited in the 
material in front of and in-between calorimeters. 
The response for hadronic jets~\cite{ATLAS-CONF-2011-032}
is $\sim 50\%$ of the true energy 
for $\ptJet=20\,\GeV$ and $|\etaJet|<0.8$ 
and rises both with \ptJet\ and \etaJet.  For $|\etaJet|<0.8$, the
response at $\ptJet=1\,\TeV$ is $\sim 80\% $.

The ATLAS trigger consists of three levels of event selection: 
Level-1 (L1), Level-2 (L2), and Event Filter. The L2 and event filter 
together form the High-Level Trigger (HLT).
The L1 trigger is implemented using custom-made electronics, while the 
HLT is based on fast data reconstruction online algorithms running 
on commercially available computers and networking systems. 
The triggers relevant for this  analysis are 
the L1 minimum bias triggers (MBTS) and the L1 and HLT calorimeter triggers.
The minimum bias trigger is based on signals from 32 scintillation
counters located at pseudorapidities  $2.09 <|\eta| < 3.84$.  
Because non-diffractive events fire the MBTS with
high efficiency and negligible bias,  this trigger can be used to study
jets with low \ptJet.  However, MBTS triggers were highly prescaled at large
instantaneous luminosities, making them unsuitable for studies 
of high \pt\ jets that are produced at low rate.
A series of single jet inclusive
triggers with different jet $E_T$ thresholds and prescales were
deployed to ensure that significant data samples were taken over the full
range of \ptJet~\cite{bib:jet}.   
\section{Monte Carlo Samples}
\label{sec:samples}

Several Monte Carlo samples are used in this analysis. Some
 samples were processed
with the 
ATLAS full detector  simulation~\cite{bib:atlas-sim} which is based on the 
{\sc GEANT4} toolkit~\cite{bib:g4}. The simulated events are then
passed through the same reconstruction software as the data.
These are used to model the response of the
detector and to correct the data for experimental effects.
The baseline Monte Carlo sample used to determine these
corrections is produced using \Pythia\cite{Sjostrand:2006za} 6.421  with the 
ATLAS tune~{\sc AMBT1}  which uses the MRST2007LO* PDFs~\cite{Sherstnev:2007nd}
and was derived using the measured properties of
minimum bias events~\cite{Aad:2010ir}. Several other 
fully simulated samples are used
to assess systematic uncertainties: \Pythia\ using the \Perugiatwenty\
tune~\cite{Skands:2010ak} (CTEQ5L PDFs~\cite{Pumplin:2002vw}); 
Herwig 6.5~\cite{Corcella:2000bw} using Jimmy 3.41~\cite{Butterworth:1996zw}
and \Herwigpp\ 2.4.2~\cite{Bahr:2008pv}   (MRST2007LO* PDFs).

Additional Monte Carlo generator samples 
are used to
compare with the final corrected data: \Pythia 6.421  with the ATLAS MC09
tune~\cite{ATL-PHYS-PUB-2010-002} (MRST2007LO* PDFs),
 \Herwigpp\ 2.5.1~\cite{Gieseke:2011na} (MRST2007LO* PDFs) , 
Sherpa~\cite{Gleisberg:2008ta} (CTEQ6L\cite{Pumplin:2005rh}  PDFs) and 
\Pythia8 (8.105)\cite{Sjostrand:2007gs}\\ 
(MRST2007LO* PDFs).
 \section{Reconstruction and Event Selection}
\label{sec:selection}
\begin{figure*}
  \begin{center}
   \includegraphics[width=0.49\textwidth]{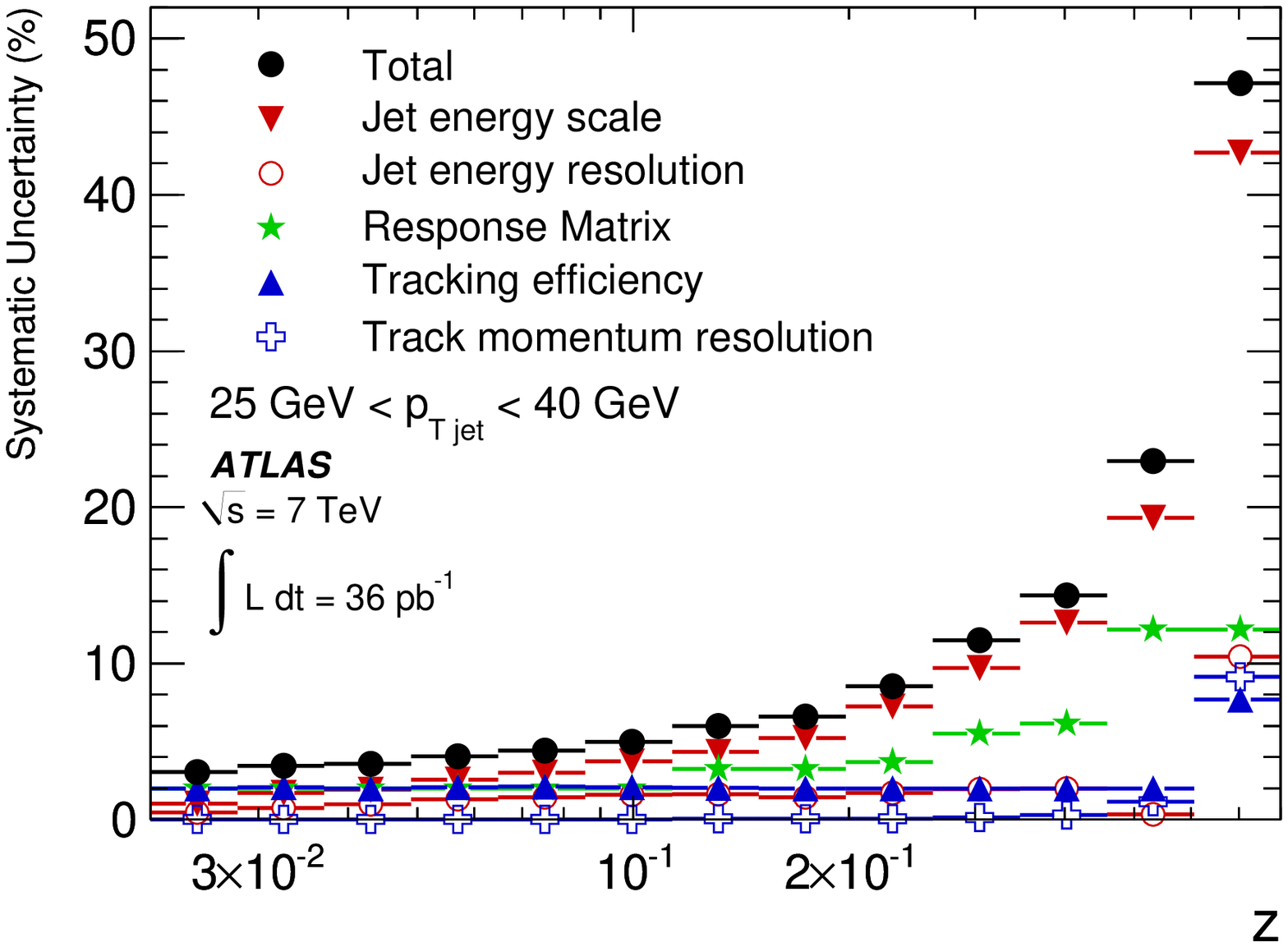}
   \includegraphics[width=0.49\textwidth]{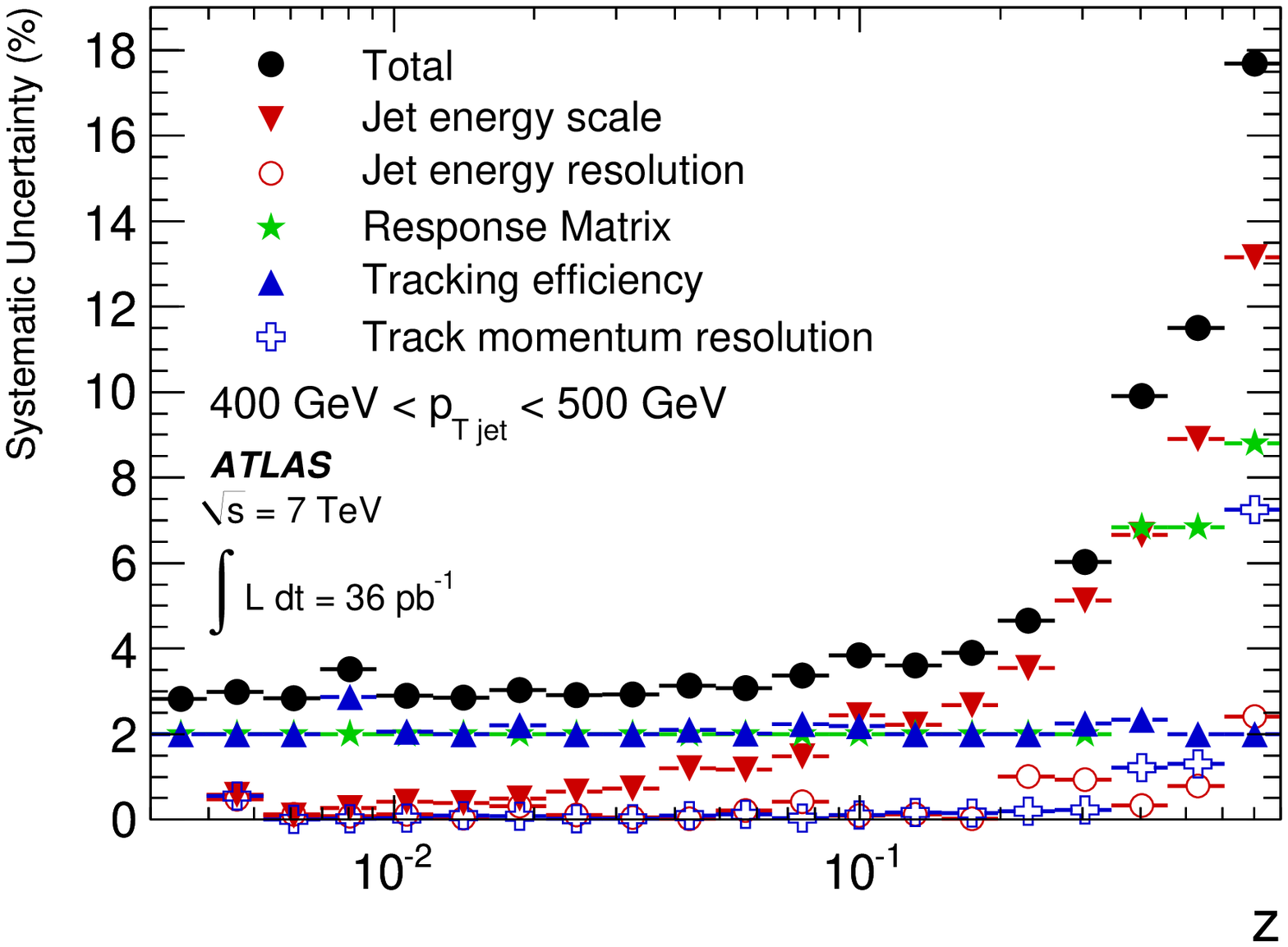}
  \end{center}
    \caption{Systematic uncertainty in $F(z,\ptJet)$ from  uncertainties 
     in  the jet energy scale and resolution, the track reconstruction efficiency and momentum
      resolution and the response matrix for $25\ \GeV <\ptJet < 40\ \GeV$ (\emph{left}) and $400\ \GeV <
     \ptJet < 500\ \GeV$ (\emph{right}).  The total uncertainty from the combination is also shown.}
  \label{Fig:UnfoldingSystProjections}
\end{figure*}

\begin{figure*}
  \begin{center}
   \includegraphics[width=0.49\textwidth]{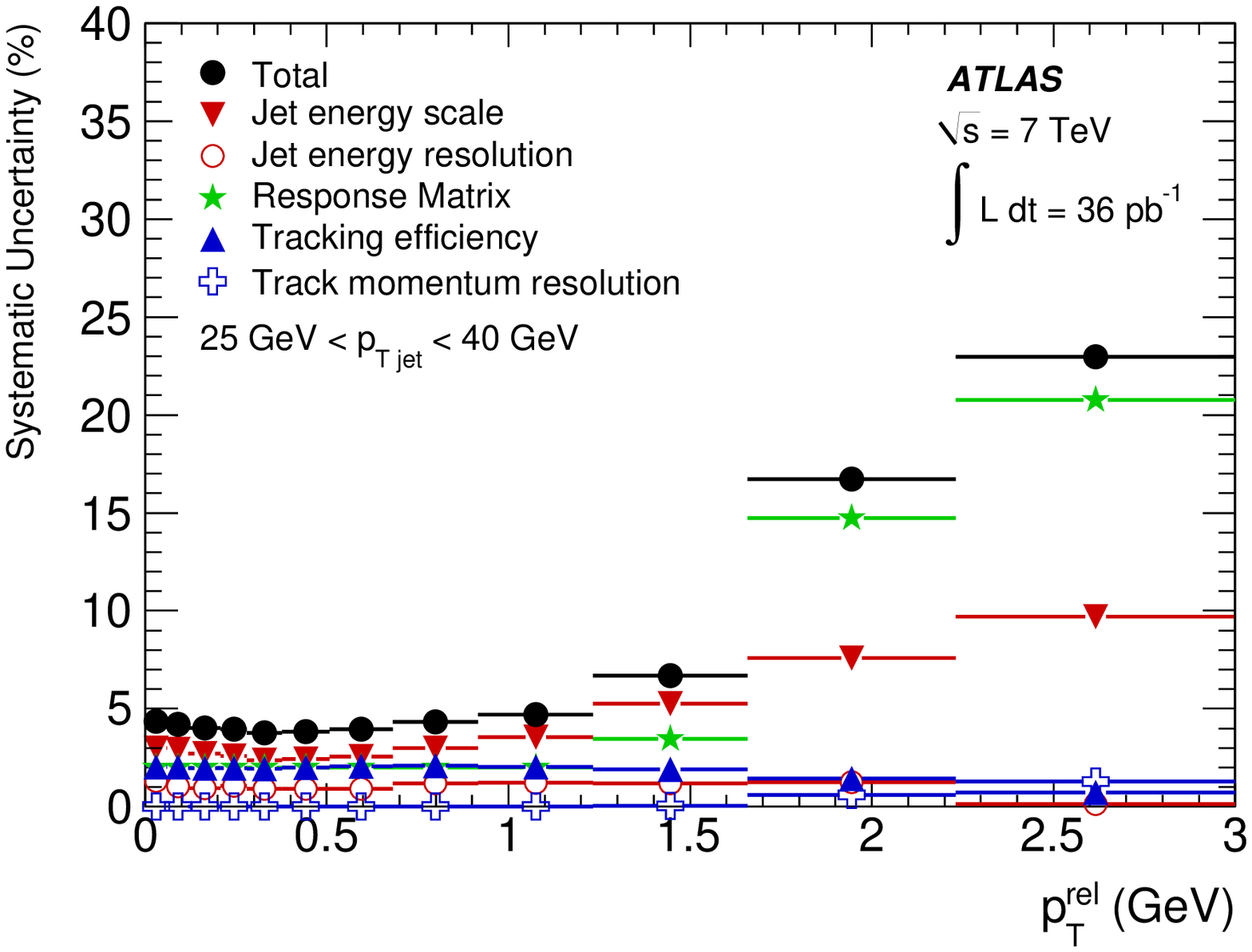}
   \includegraphics[width=0.49\textwidth]{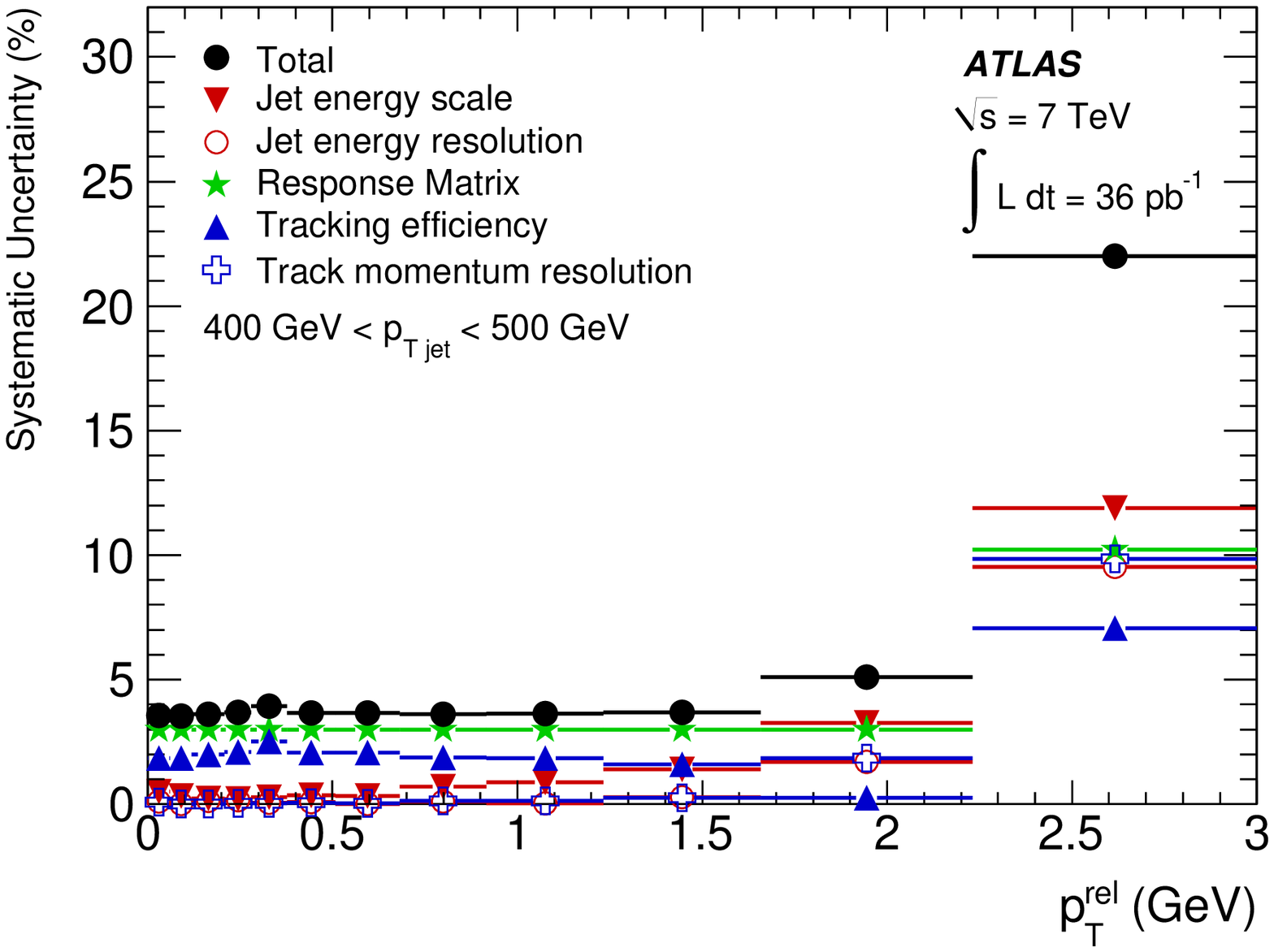}
  \end{center}
    \caption{Systematic uncertainty  in $f(\ptrel,\ptJet)$ from  
      uncertainties in the jet energy scale and resolution, the track reconstruction efficiency and momentum
      resolution and the response matrix for $25\ \GeV <\ptJet < 40\ \GeV$ (\emph{left}) and $400\ \GeV <
     \ptJet < 500\ \GeV$ (\emph{right}).  The total uncertainty from the combination is also shown.  }
  \label{Fig:UnfoldingPtRelSystProjections}
\end{figure*}

\begin{figure*}
  \begin{center}
   \includegraphics[width=0.49\textwidth]{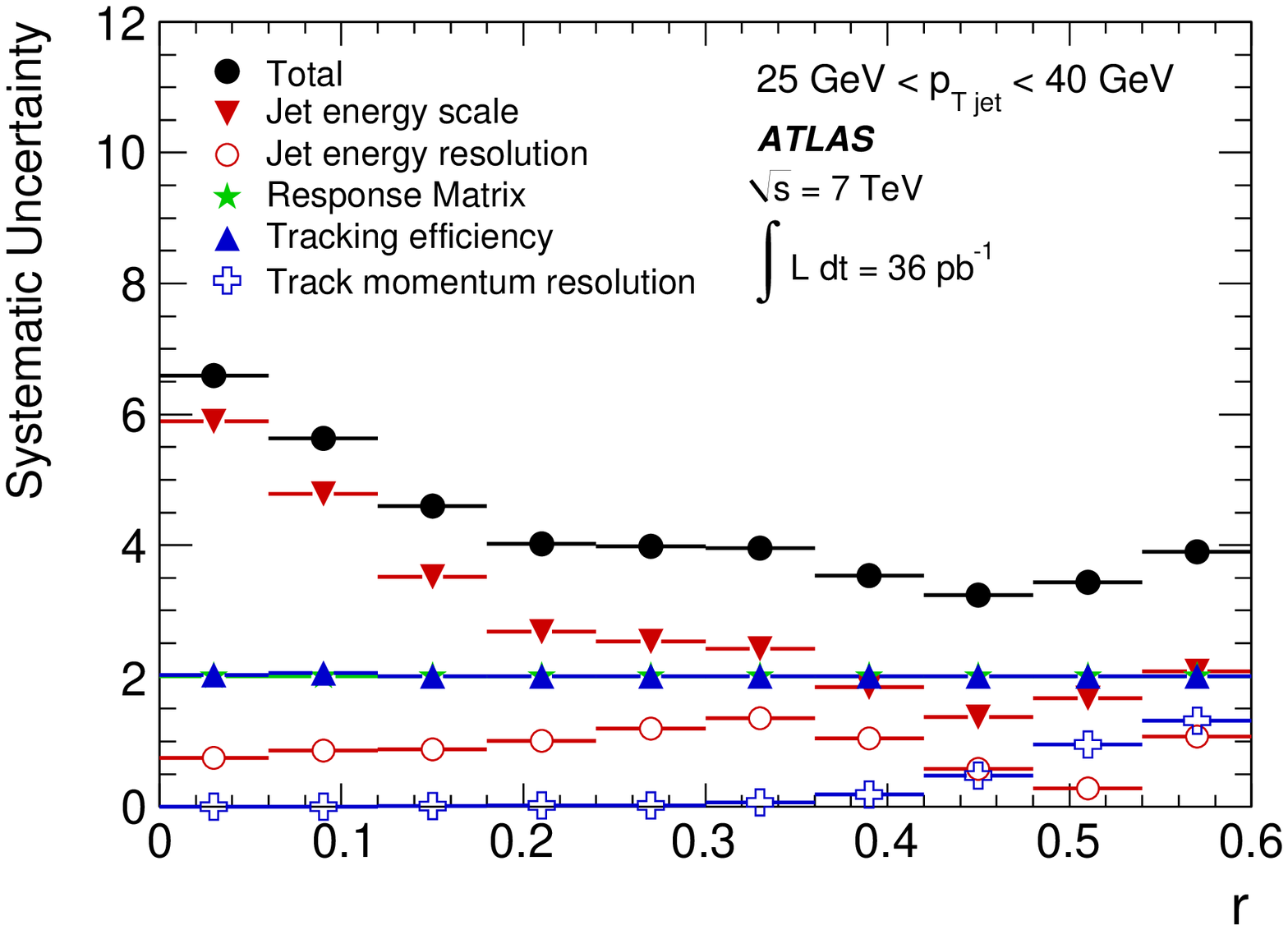}
   \includegraphics[width=0.49\textwidth]{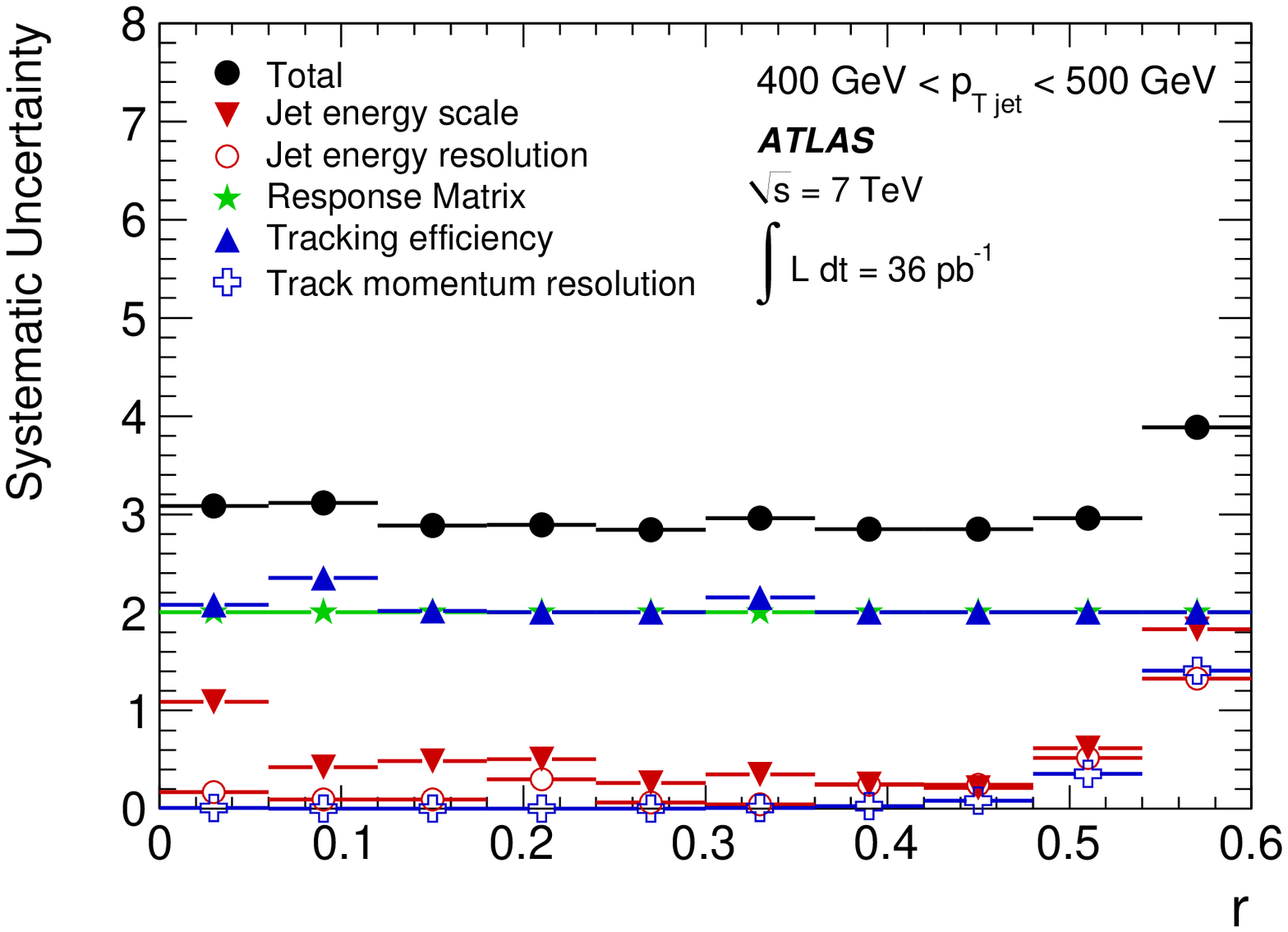}
  \end{center}
    \caption{Systematic uncertainty  in  $\rho_{ch}(r,\ptJet)$ from 
      uncertainties  in the jet energy scale and resolution, the track reconstruction efficiency and momentum
      resolution and the response matrix for $25\ \GeV <\ptJet < 40\ \GeV$ (\emph{left}) and $400\ \GeV <
     \ptJet < 500\ \GeV$ (\emph{right}). The total uncertainty from the combination is also shown.}
\label{Fig:UnfoldingRhoSystProjections}
\end{figure*}
\begin{figure*}
  \begin{center}
   \includegraphics[width=0.49\textwidth]{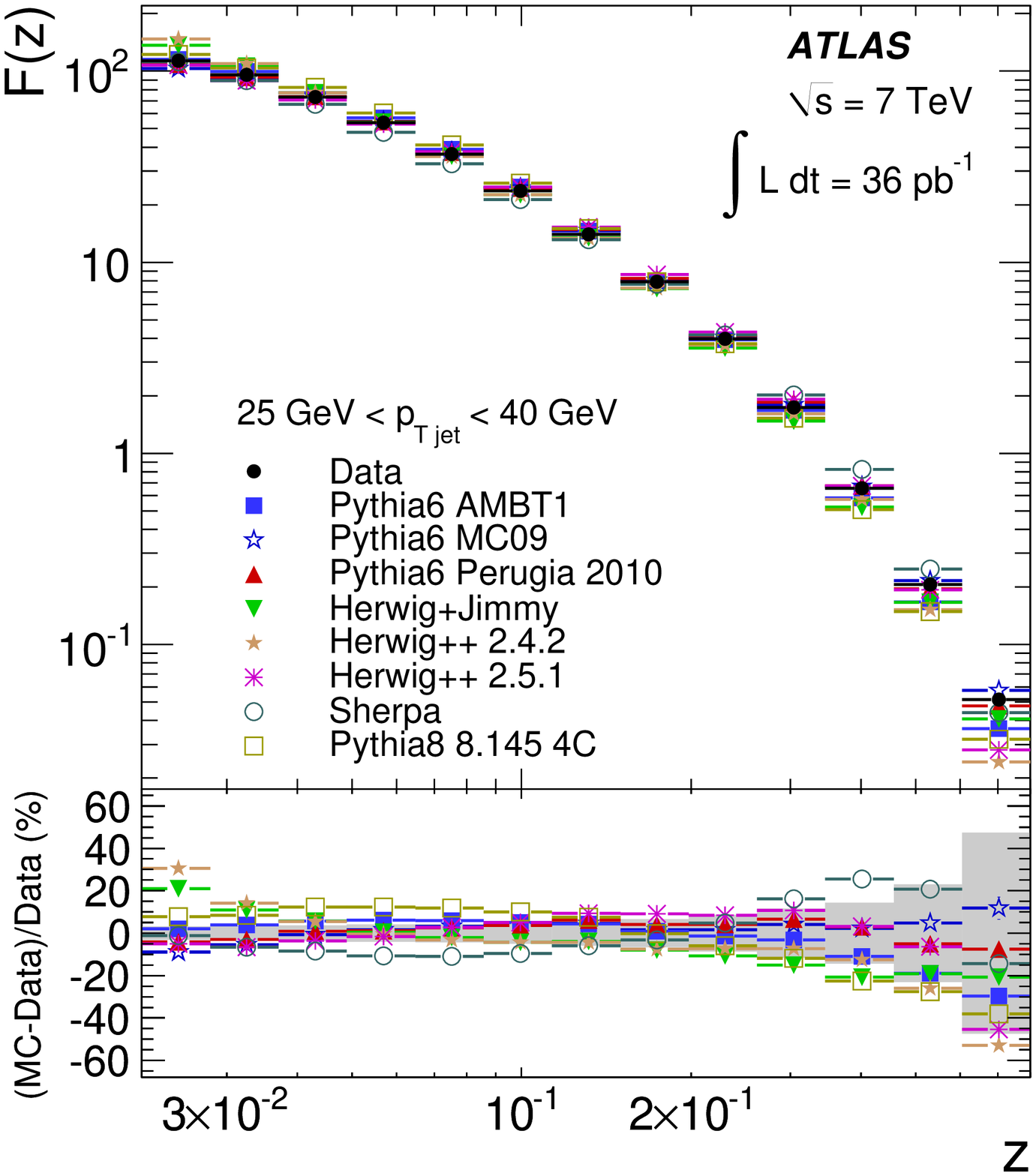}
   \includegraphics[width=0.49\textwidth]{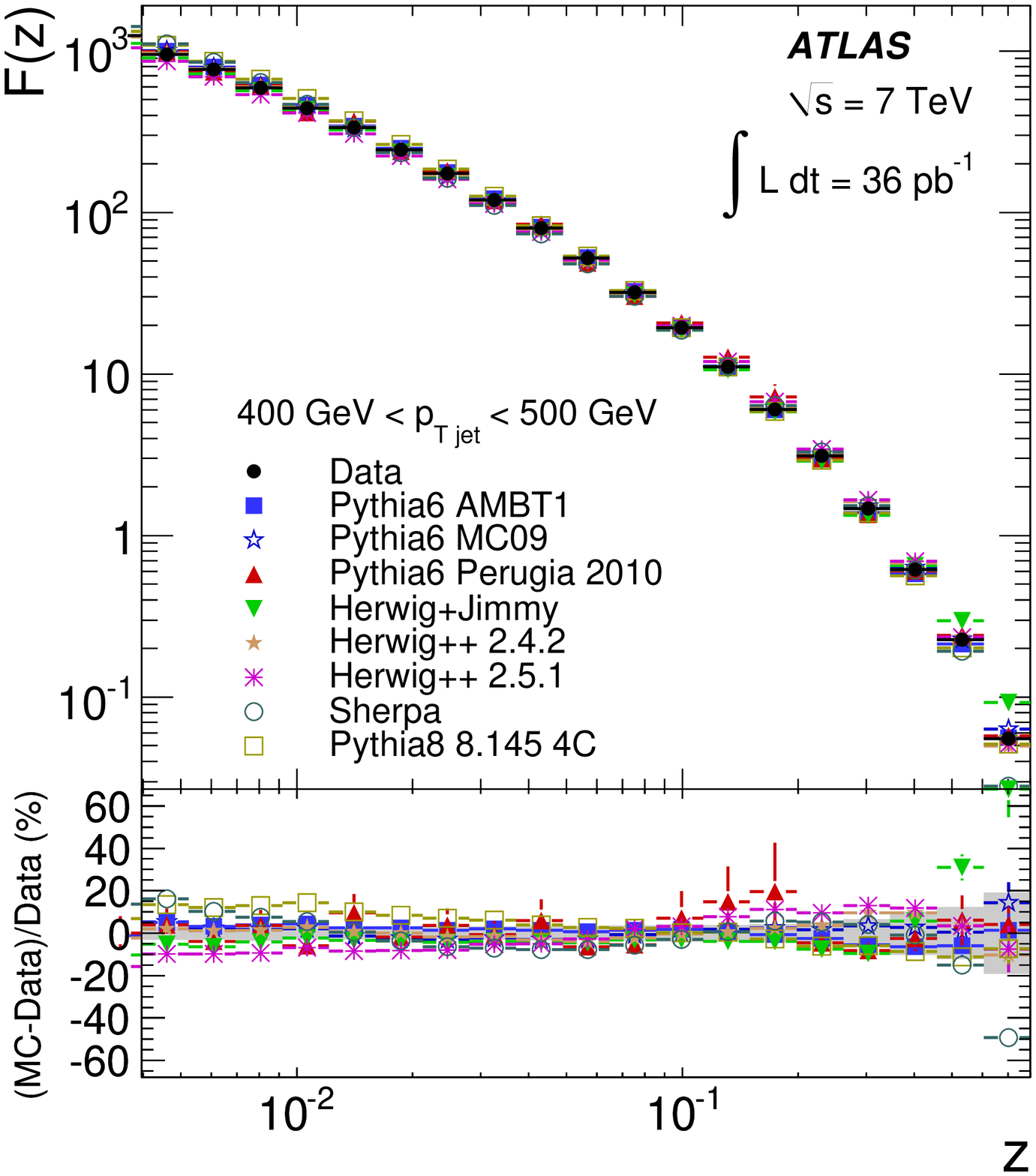}
  \end{center}
    \caption{Distributions of $F(z)$ for $25\ \GeV < \ptJet < 40\ \GeV$
      (\emph{left}) and $400\ \GeV < \ptJet < 500\ \GeV$ (\emph{right}). The gray band indicates the total
      uncertainty.}
  \label{Fig:UnfoldingDataProjections}
\end{figure*}

Events are required to have at least one primary vertex reconstructed
using ID tracks.  If the event has multiple primary vertices, the vertex
with the largest $\sum (\ptTrack)^2$
is tagged as the 
hard-scattering vertex.

Jets are reconstructed using the
infrared- and collinear-safe \antikt\ algorithm~\cite{bib:antikt2} 
with radius parameter $R_c=0.6$ 
using the FastJet package~\cite{bib:antikt3}.
The detector input is based on topological clusters~\cite{bib:topo}.
A topological cluster is defined to have an energy equal to the 
energy sum of all the included calorimeter cells, zero mass and a 
reconstructed direction calculated from the weighted averages 
of the pseudo-rapidities and azimuthal angles of the constituent 
cells. The weight used is the absolute cell energy and the 
positions of the cells are relative to the nominal ATLAS coordinate 
system.
The energy of these clusters is measured at the electromagnetic scale,
which provides the appropriate calibration for electrons and photons.
A \ptJet\ and \etaJet\ dependent
calibration is then applied to each jet\cite{ATLAS-CONF-2011-032}.  
These calibrations are  
based on comparing the
response from simulated calorimeter jets to that of 
jets reconstructed using generator particles and matched to the
reconstructed jets in $\eta$-$\phi$  space. 
The $\eta$-$\phi$ position of the jet (and hence its momentum)
is corrected to account for the fact that the primary vertex of the interaction
is not at the geometric centre of the detector. 
Quality criteria are applied to ensure that
jets are not produced by noisy calorimeter cells, and to
avoid problematic detector regions. The jet energy is corrected
for the presence of additional $pp$ interactions in the
same bunch crossing using correction constants measured {\it in-situ}
that depend on the number of reconstructed primary 
vertices.

Jets are required to have $\abs{\etaJet}<1.2$. 
For events selected with the MBTS trigger, jets are required to pass
a minimum cut of $\ptJet > 20$~GeV.  
For events selected using jet triggers,
a trigger-dependent minimum \ptJet\ threshold is imposed 
on jets used in the final measurements  to ensure a jet trigger efficiency larger than 99\%.

Tracks are selected using the following cuts:
\begin{eqnarray*}
&\ptTrack >0.5\,\GeV, \ N_{\rm pixel} \ge 1, \ N_{\rm SCT} \ge 6, & \\
&|d_0| < 1.5\,{\rm mm}, \ |z_0\sin\theta| < 1.5\,{\rm mm},& 
\end{eqnarray*}
where $N_{\rm pixel}$ and $N_{\rm SCT}$ are the number of hits from the
pixel and SCT detectors, respectively, that are associated with the 
track and
$d_0$ and $z_0$ are the transverse and longitudinal impact parameters 
measured with 
respect to the hard-scattering vertex.

Tracks are associated with jets using a simple geometric algorithm.  If
the distance in $\eta$-$\phi$ between the track and the jet
is less than the radius parameter used in the jet reconstruction ($R_c=0.6$), 
the tracks are considered to belong to the jet.  Track parameters
are evaluated at the perigee to the primary vertex
and are {\it not} extrapolated to the 
calorimeter.  
This simple association
algorithm facilitates comparison with particles from the event generator
whose parameters correspond
to those measured at the primary vertex..
\section{Analysis}
\label{sec:analysis}

The results presented here are obtained using four measured distributions:
the jet transverse momentum spectrum, 
${dN_{\mathrm jet}(\ptJet)}/{d\ptJet}$, and
three differential distributions of the number of charged tracks,
${dN_{\mathrm tracks}(z,\ptJet)}/{dz}$, 
${dN_{\mathrm tracks}(\ptrel,\ptJet)}/{d\ptrel}$ and
${dN_{\mathrm tracks}(r,\ptJet)}/{dr}$.
To facilitate comparison with the predictions of Monte Carlo event generators,
these distributions are corrected for detector acceptance, reconstruction
efficiency and migration due to track and jet momentum resolution effects.
This correction procedure is called unfolding.
The distributions $F(z,\ptJet)$, $f(\ptrel,\ptJet)$ and $\rho_{ch}(r,\ptJet)$
are obtained from the charged particle differential distributions
by normalizing the distribution for each \ptJet\ range to the value
of $N_{\mathrm jet}(\ptJet)$ obtained from the unfolding of the jet 
transverse momentum
spectrum.   
This paper presents results for $\ptJet > 25$~GeV; however,
to decrease the systematic uncertainty associated with the modeling
of the \ptJet\ spectrum, 
jets with $20\ \GeV < \ptJet < 25\ \GeV$ are also used in the unfolding.

A Bayesian iterative unfolding method \cite{D'Agostini:1994zf}
implemented in the RooUnfold \cite{RooUnfold} software package is used. 
This procedure takes as its input the measured distributions and a
response matrix obtained from simulated data that provides a mapping
between reconstructed objects and those obtained directly from the
event generator.
This response matrix is not unitary because in mapping from generator to
reconstruction some events and objects are lost due to inefficiencies
and some are gained due to misreconstruction or migration of truth
objects from outside the fiducial acceptance into the reconstructed 
observables.  It is therefore not possible to obtain the unfolded distributions
by inverting the response matrix and applying it to the measured data.
Instead, an
assumed  truth distribution (the ``prior'')
is selected, the response matrix is applied and 
the resulting trial reconstruction set is compared to the observed 
reconstruction set. 
A new prior is then constructed from the old prior
and the difference between the trial and the observed distributions.
The procedure can iterated until this difference becomes small.
Monte Carlo based studies of the performance of the procedure demonstrate
that in this analysis no iteration is necessary.
The initial truth prior is taken to be the prediction of the  baseline 
Monte Carlo generator.  Systematic uncertainties associated with this
choice and with the modeling of the response matrix are discussed in 
Section~\ref{sec:systematics}.

\begin{figure}
  \begin{center}
   \includegraphics[width=0.48\textwidth]{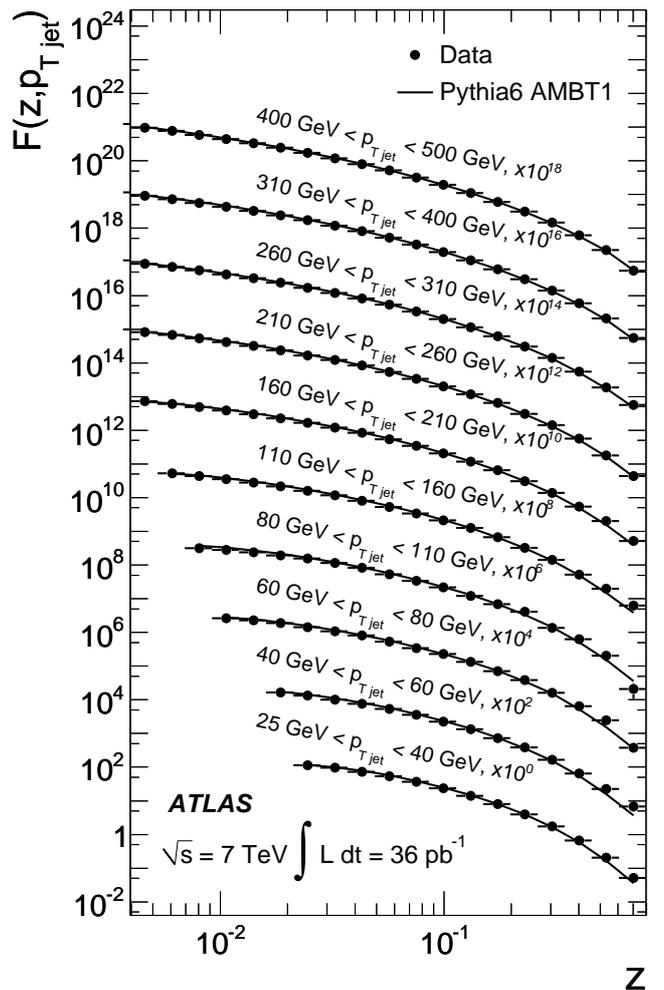}
  \end{center}
    \caption{Distributions of $F(z)$ in bins of $\ptJet$.  The circles show unfolded data and the
lines are the predictions from AMBT1 \Pythia. }
  \label{Fig:zProjections}
\end{figure}
\section{Systematic Uncertainties}
\label{sec:systematics}

The following sources of systematic uncertainties are considered:
\begin{enumerate}
\item The jet energy scale (JES) and resolution (JER) uncertainties which 
affect the measurement of the number of jets in a given \ptJet\ bin and consequently the measured value of 
$z$.
\item The track reconstruction efficiency and momentum reconstruction uncertainties which affect the number of tracks in each 
$z$, \ptrel\ and $N_{ch}(r)$ bin.
\item The uncertainty in the response matrix which is derived using a 
particular Monte Carlo sample and 
depends on the details of the event generator.
\item Potential bias due to the failure of the unfolding procedure to converge
to the correct value.
\end{enumerate}
These systematic uncertainties are addressed using Monte Carlo methods.

The first two systematic uncertainties, potential bias due to incorrect
Monte Carlo modeling of the JES and/or JER
and potential bias due to mismodeling by the simulation
of the track reconstruction efficiency and/or resolution,
are studied by modifying the detector response in simulated
data.
These modified Monte Carlo events are then unfolded and
compared to the baseline.
The systematic uncertainty on the JES is studied by varying
the jet energy response by its uncertainty.  The JES uncertainty
varies from 4.6\%\ at $\ptJet=20 $~GeV to ~2.5\%\ at
$\ptJet=500 $~GeV\cite{ATLAS-CONF-2011-032}.
Systematic uncertainties on the JER are studied by broadening
the jet energy resolution with an additional \etaJet\ and \ptJet\ 
dependent Gaussian term.
The uncertainty on the JER is  below 14\%\ for the full
\ptJet\ and \etaJet\ range used in this analysis\cite{ATLAS-CONF-2010-054}.
The uncertainty on the tracking efficiency is studied by
randomly removing a fraction of the tracks in the simulated data.
Uncertainties on the tracking efficiency are $\eta$-dependent and vary
between 2\%\ and 3\%\ for the relevant range of $\eta^{\mathrm track}$\cite{mb2paper}, dominated  by the
accuracy of the 
description of the detector material in the simulation.
In addition, there can be a loss of tracking efficiency in the core
of jets at high \ptJet\ due to a single pixel hit receiving contributions from
more than track. Studies of such hit sharing  show that 
the simulation and data agree well and that 
the resulting systematic uncertainty is negligible for \ptJet$<500\ \GeV$.
Uncertainties on the track momentum resolution are parametrized
as an additional $\eta$-dependent broadening of the resolution in curvature 
with values that vary from $0.0004$~GeV$^{-1}$ to 
$0.0009$~GeV$^{-1}$\cite{ATLAS-CONF-2011-003}.

While the studies described above account for systematic uncertainties 
associated with the accuracy of the detector simulation, they do not
account for the fact that the response matrix itself depends on the
fragmentation properties of the jets  and hence on the physics description in
the event generator.  Because the response of the calorimeter to hadrons
depends on the hadron momentum\cite{ATLAS-CONF-2011-028}, the JES 
depends at the few per cent level on the momentum spectrum of
particles within the jet.  Because the probability that a track will
share hits in the ID with another track is dependent upon the local
density of particles within the jet, the tracking resolution depends weakly
on the transverse profile of particles within the jet.  
These effects have been studied by unfolding fully simulated 
Monte Carlo samples 
created from  \Perugiatwenty, Herwig 6.5 (with Jimmy 3.41) and \Herwigpp\
using the baseline response matrix obtained with \Pythia\ AMBT1.
Differences between the unfolded results for each tune
and the true distributions obtained from that same tune are studied
as a function of $z$, \ptrel\ and $N_{ch}(r)$ for each bin in true \ptJet\
and used to assess the systematic uncertainty.

Potential bias in the unfolding procedure itself is studied by creating 
1000 pseudo-experiments where the ``data'' are drawn from the baseline
fully simulated Monte Carlo samples  
via a bootstrap method\cite{erfon:bootstrap}
and unfolding these "data"  using
the standard procedure.  The mean results obtained from these samples show
negligible bias and have a spread that is consistent with the reported 
statistical uncertainties.  
The systematic uncertainty due to the unfolding
procedure is thus deemed to be negligible in comparison to the other
uncertainties.

The resulting systematic uncertainties on 
$F(z,\ptJet)$, $f(\ptrel,\ptJet)$ and 
$\rho_{ch}(r,\ptJet)$ for the $25\ \GeV < \ptJet < 40\ \GeV$ 
(\emph{left}) and $400\ \GeV < \ptJet < 500\ \GeV$ (\emph{right})
are shown in Figures~\ref{Fig:UnfoldingSystProjections} 
through~\ref{Fig:UnfoldingRhoSystProjections}.  For 
$F(z,\ptJet)$, uncertainties on the tracking efficiency and response
matrix dominate at low $z$ while the jet energy scale dominates at
high $z$.  For $f(\ptrel,\ptJet)$ the jet energy scale, response matrix
and tracking efficiency uncertainties are all significant and the overall
uncertainty rises with \ptrel.
For $\rho_{ch}(r,\ptJet)$, the response matrix and tracking efficiency
uncertainties are significant for all \ptJet\ and $r$ while the jet energy
scale contribution is most important for small \ptJet.
\section{Results}
\label{sec:results}

\begin{figure*}
    \begin{center}
   \includegraphics[width=0.48\textwidth]{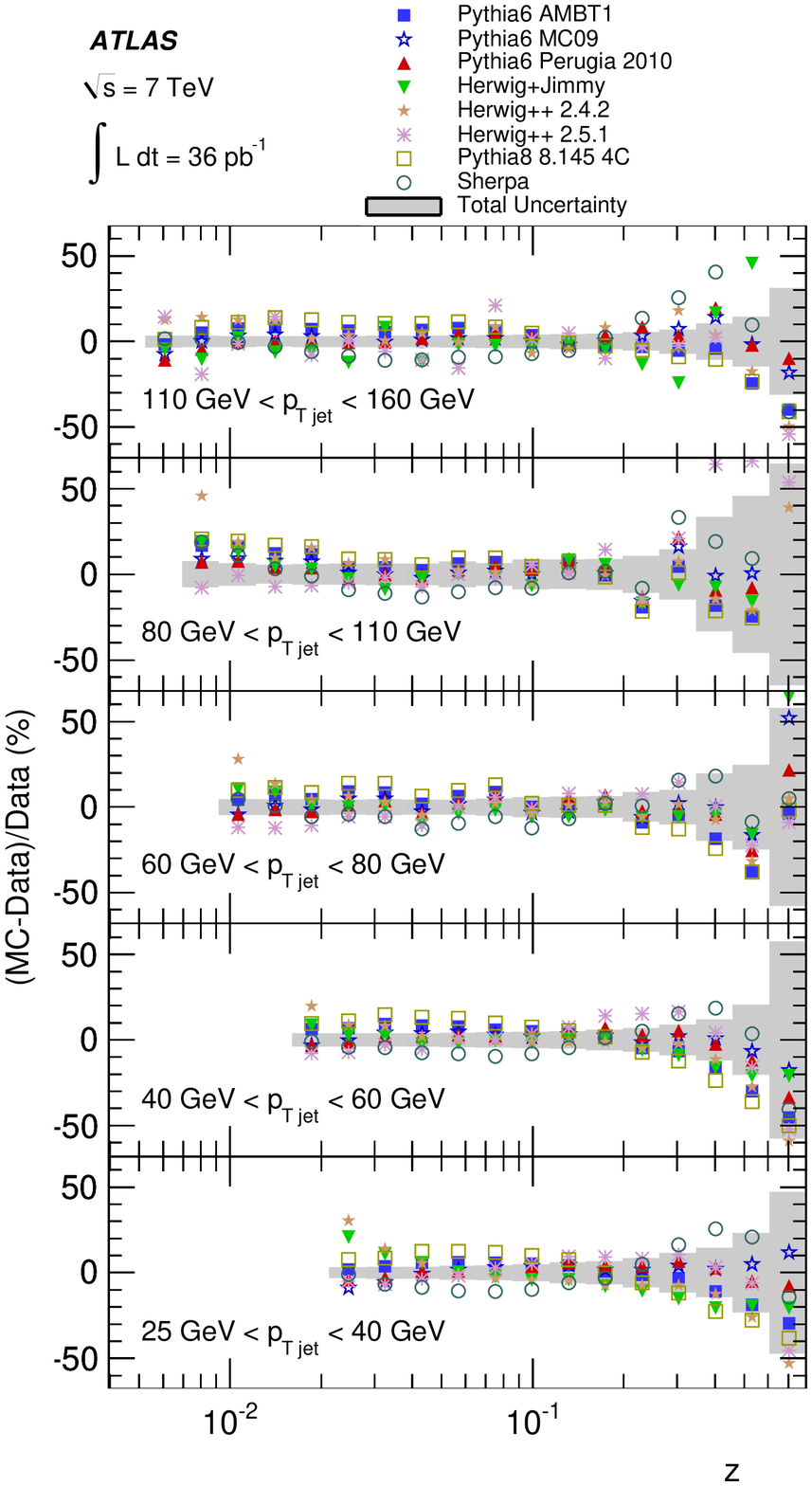}
   \includegraphics[width=0.48\textwidth]{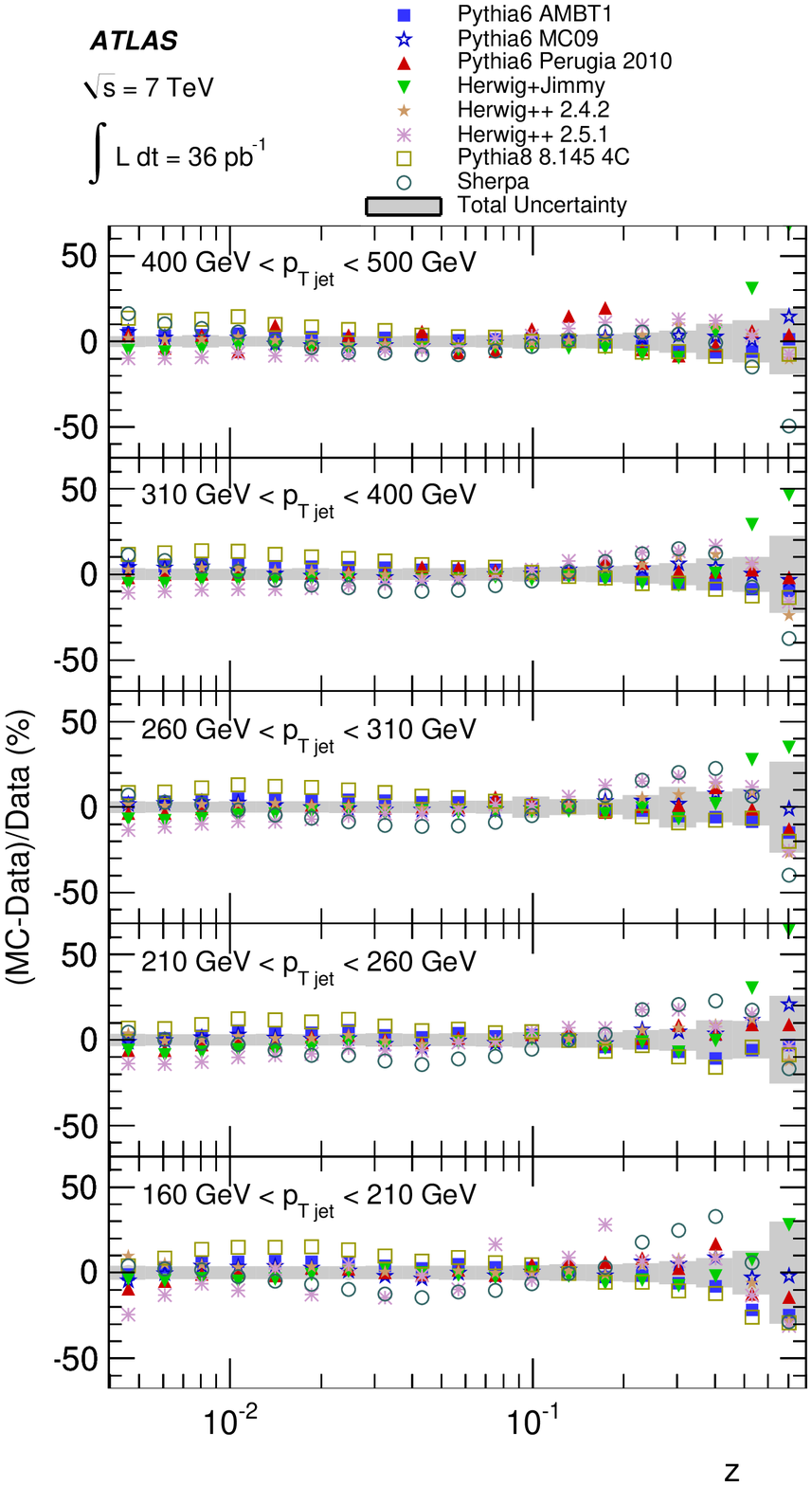}
    \end{center}
\caption{The ratio of $F(z)$ predicted by various Monte Carlo generators
to that measured.
The gray band indicates the combined statistical and systematic
uncertainties.}
\label{Fig:zTruthOver0}
\end{figure*}

This section presents comparisons of  acceptance-corrected, unfolded data 
to the predictions of several Monte Carlo generators.  The gray
band on all the figures indicates the total uncertainty which is 
dominated by  the systematic uncertainty.
Figure~\ref{Fig:UnfoldingDataProjections} shows distributions of $F(z)$ in 
two bins of \ptJet.
Figure~\ref{Fig:zProjections} shows distributions 
of $F(z)$ in all bins of \ptJet\ compared
to AMBT1  Monte Carlo. Comparisons of the data and the Monte Carlo samples
are shown in Figure~\ref{Fig:zTruthOver0}.
All the \Pythia~6 tunings show
good agreement with the data. Herwig+Jimmy disagrees with the data 
at large $z$ for 
 $\ptJet> 200\ \GeV $. \Herwigpp~2.5.1 is below the data at low $z$ for
 $\ptJet> 100\ \GeV $ while \Herwigpp~2.4.2  has too many particles at  low $z$
for  $\ptJet< 100\  \GeV$. \Pythia 8 and Sherpa provide a poor 
description of the data.

\begin{figure*}
\subfigure{
   \includegraphics[width=0.48\textwidth]{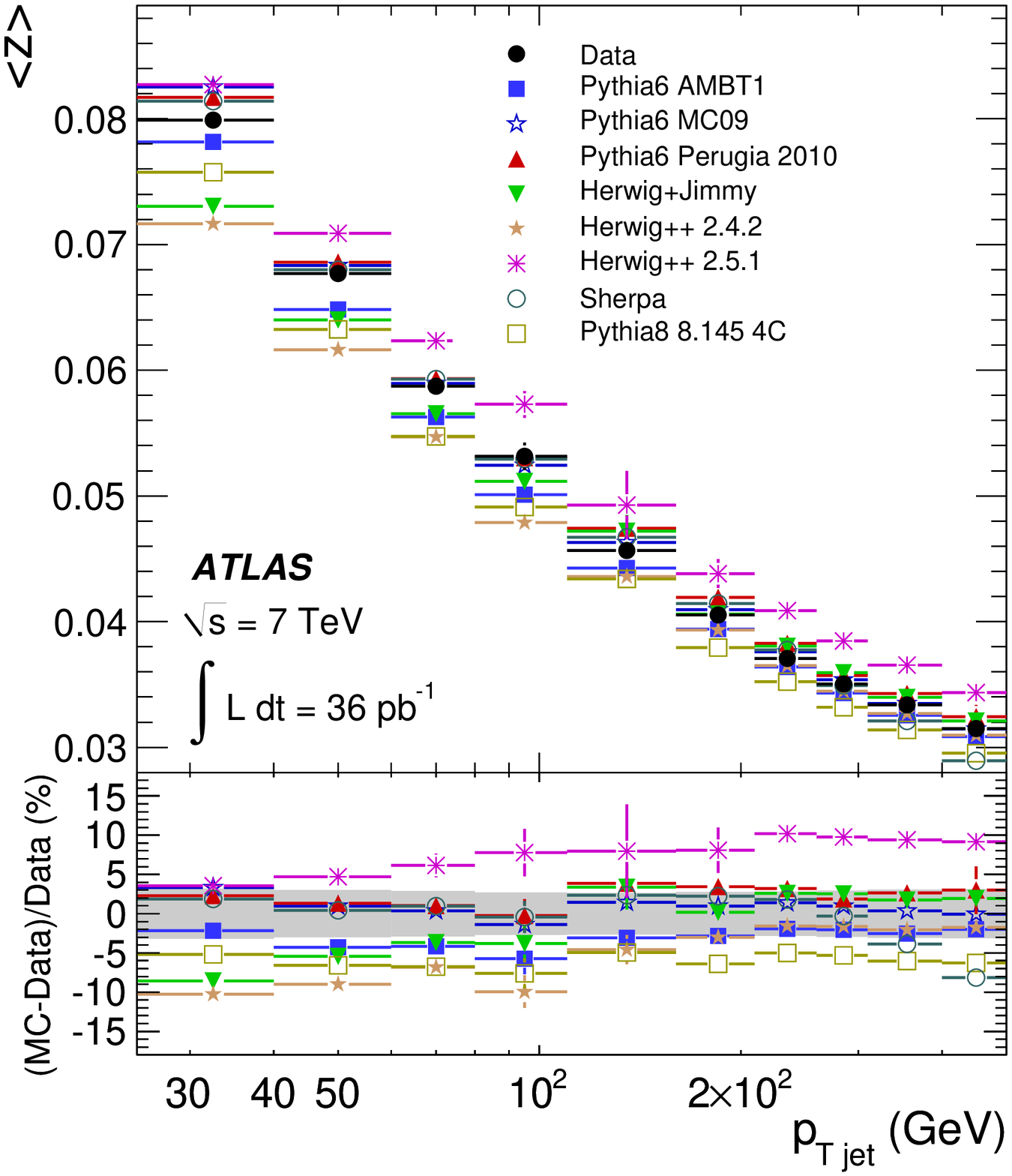}
}
\subfigure{
 \includegraphics[width=0.48\textwidth]{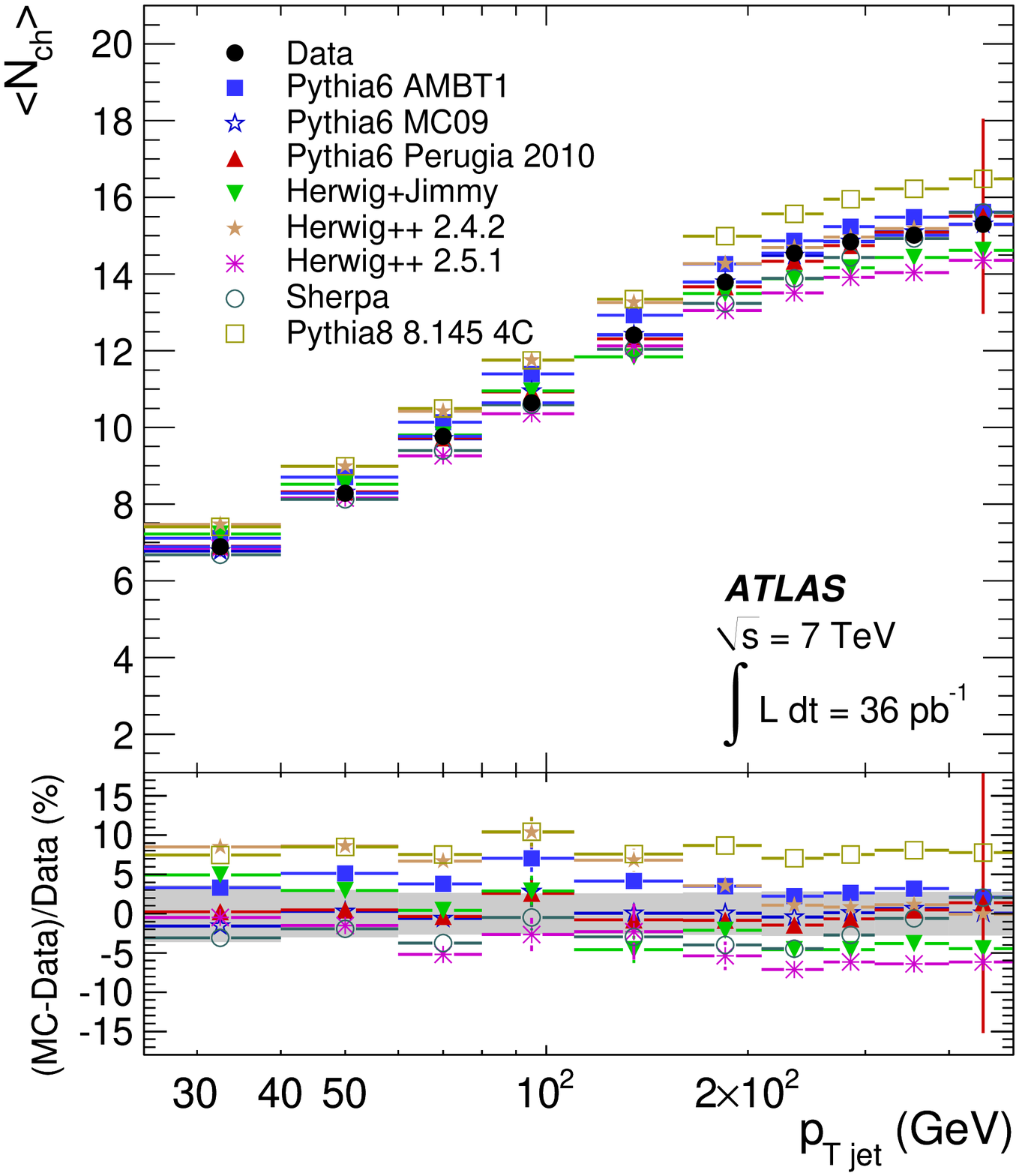}
}
    \caption{Distributions of $<z>$ (left) and of the mean number of charged particles selected with
    the requirement \ptTrack\  > 500 MeV (right) as a function of \ptJet\  for data and various
    Monte Carlos. The gray band indicated the total uncertainty.}
\label{Fig:UnfoldingData}
\label{Fig:meanmult}
\end{figure*}

Figure~\ref{Fig:UnfoldingData} (\emph{left}) shows the
distribution of $\langle z \rangle$ for the data and for a selection of
Monte Carlo samples as a function of \ptJet.  A comparison with
the Monte Carlo generators shows that  the AMBT1 and
MC09 \Pythia\
and \Perugiatwenty\
datasets show good agreement with the data 
over the entire
\ptJet\ range.  The  agreement with  Herwig+Jimmy  is satisfactory.  \Herwigpp\ 2.5.1 
is inconsistent with the data for  $\ptJet> 40\ \GeV$ and 2.4.2 is inconsistent for
 $\ptJet < 100\ \GeV$. \Pythia8 \ is $\sim 8$\%\ below the data 
at all \ptJet.
\Sherpa \ agrees well.

The charged particle multiplicity as a function of $\ptJet$ is shown in Figure~\ref{Fig:meanmult}  (\emph{right}). The \Pythia~6
tunes show reasonable agreement, with AMBT1 being higher than the others. Herwig+Jimmy has slightly 
too few particles for   $\ptJet> 200\ \GeV$. \Herwigpp\ 2.4.2 (2.5.1) has too many (few) particles
for  $\ptJet< 200\ (>300)\ \GeV$. Sherpa describes the data well 
while \Pythia8 has $\sim 8$\%\ too many particles at all \ptJet.

\begin{figure*}
  \begin{center}
   \includegraphics[width=0.49\textwidth]{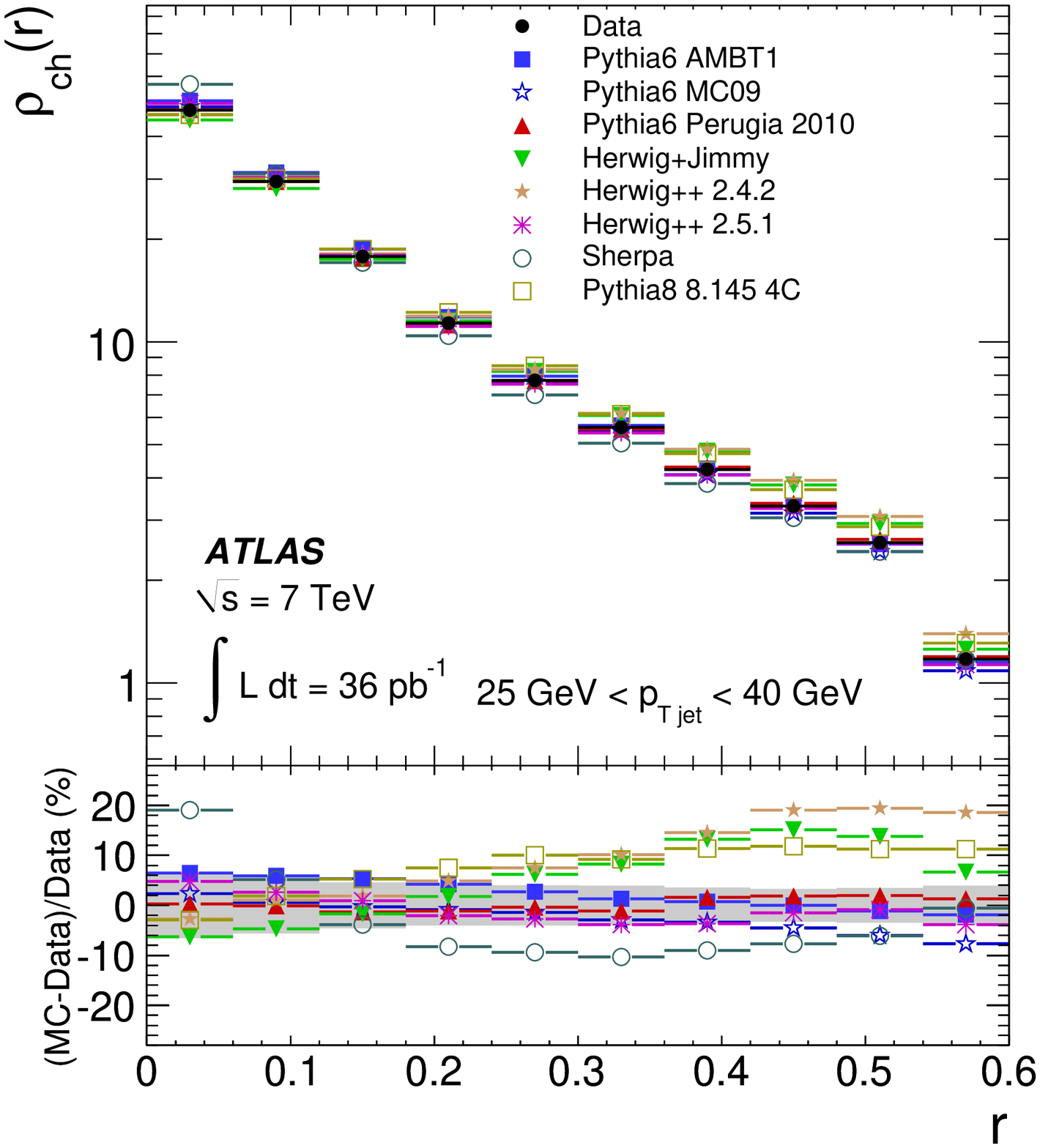}
   \includegraphics[width=0.49\textwidth]{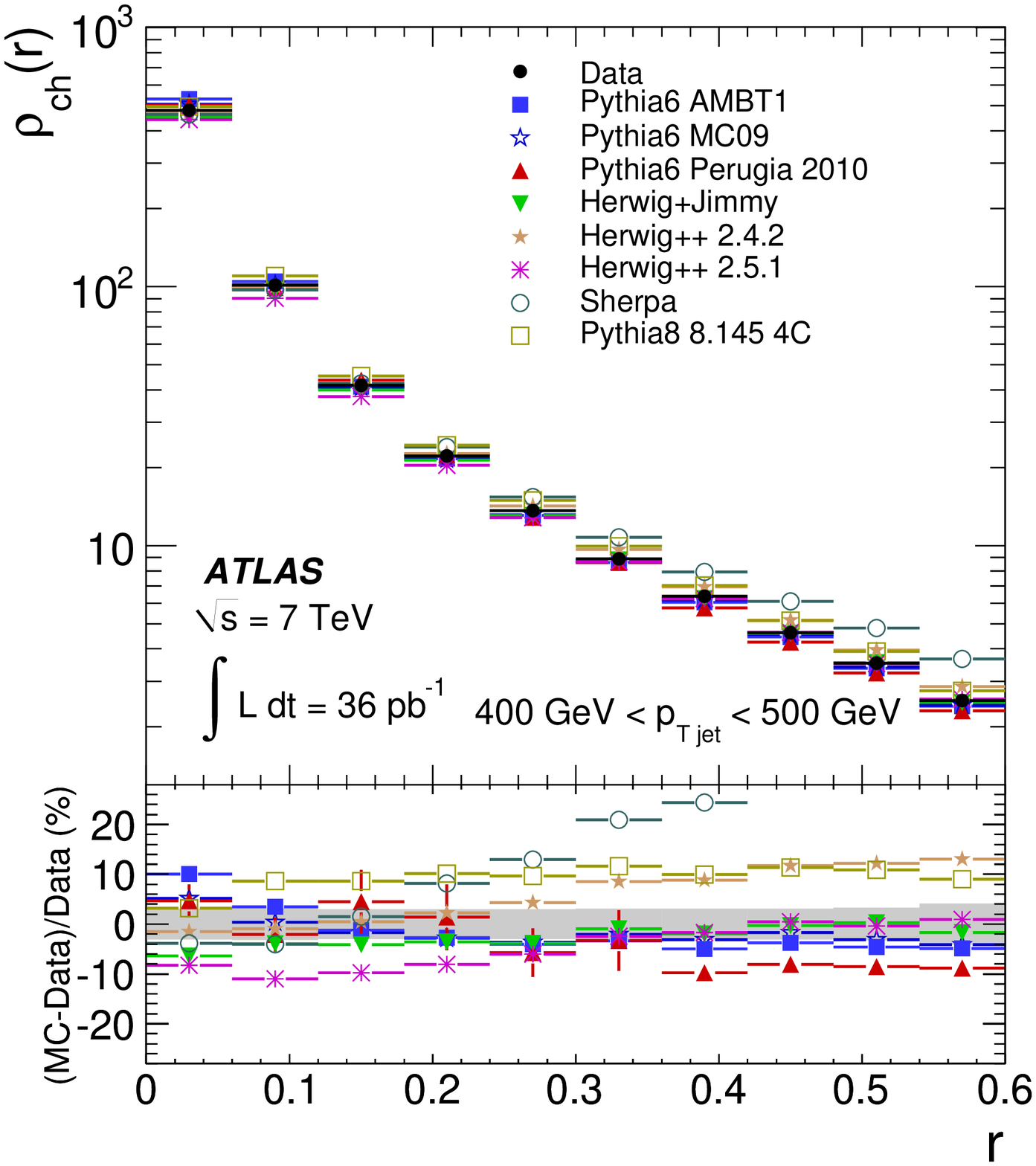}
  \end{center}
    \caption{Distributions of \rhoch\  for $25\ \GeV < \ptJet < 40\ \GeV$
      (\emph{left}) and $400\ \GeV < \ptJet < 500\ \GeV$ (\emph{right}). The gray band indicates the total
      uncertainty.}
  \label{Fig:UnfoldingDataProjectionsrho}
\end{figure*}

\begin{figure*}
  \begin{center}
   \includegraphics[width=0.49\textwidth]{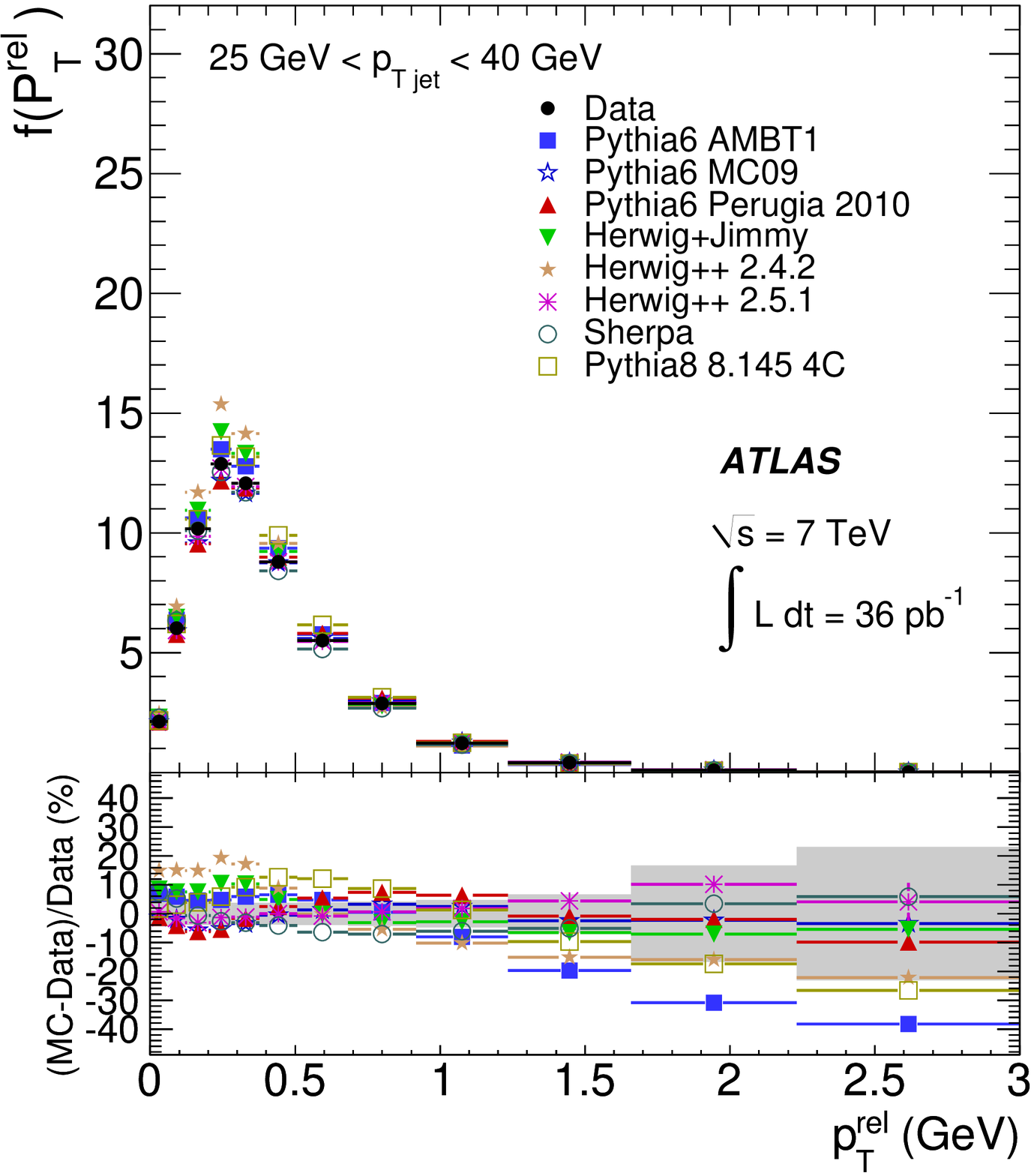}
   \includegraphics[width=0.49\textwidth]{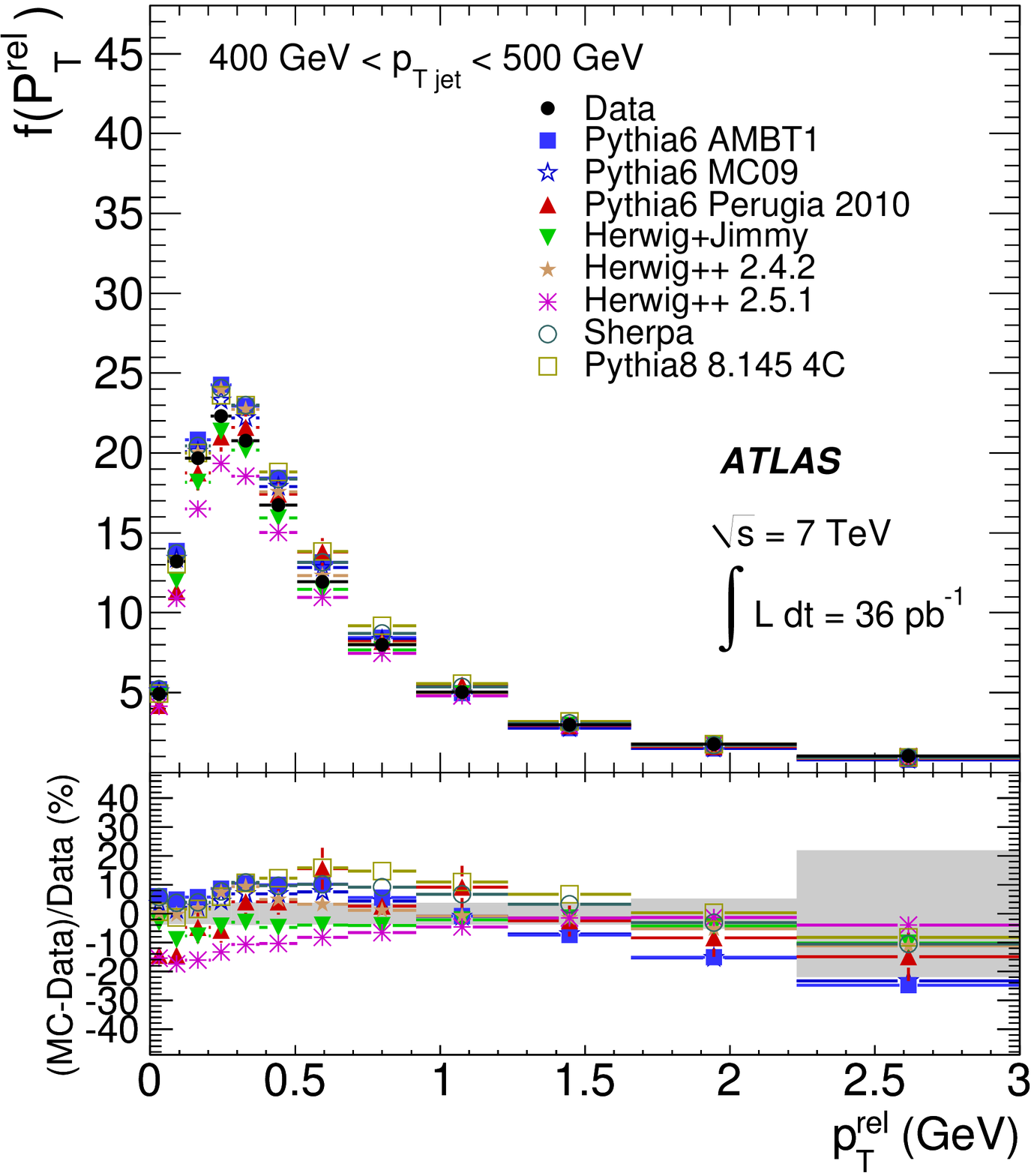}
  \end{center}
    \caption{Distributions of $f(\ptrel)$ for $25\ \GeV < \ptJet < 40\ \GeV$
      (\emph{left}) and $400\ \GeV < \ptJet < 500\ \GeV$ (\emph{right}). The gray band indicates the total
      uncertainty.}
  \label{Fig:UnfoldingDataProjectionsptrel}
\end{figure*}

\begin{figure}
  \begin{center}
   \includegraphics[width=0.48\textwidth]{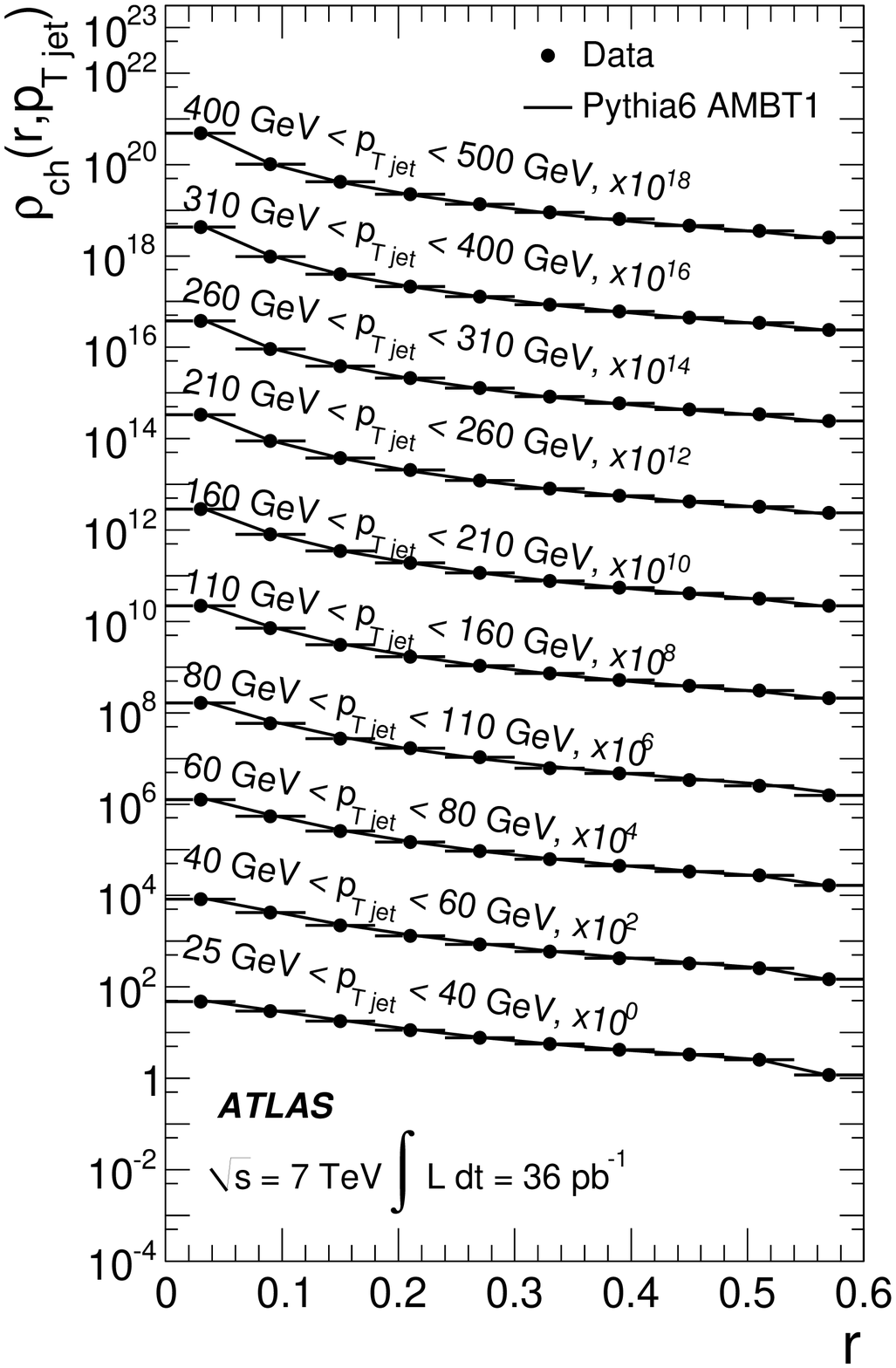}
  \end{center}
    \caption{Distributions of \rhoch. 
The circles show unfolded data.  The
lines are the predictions from AMBT1 \Pythia.
}
  \label{Fig:rhoProjections}
\end{figure}

\begin{figure}
  \begin{center}
   \includegraphics[width=0.48\textwidth]{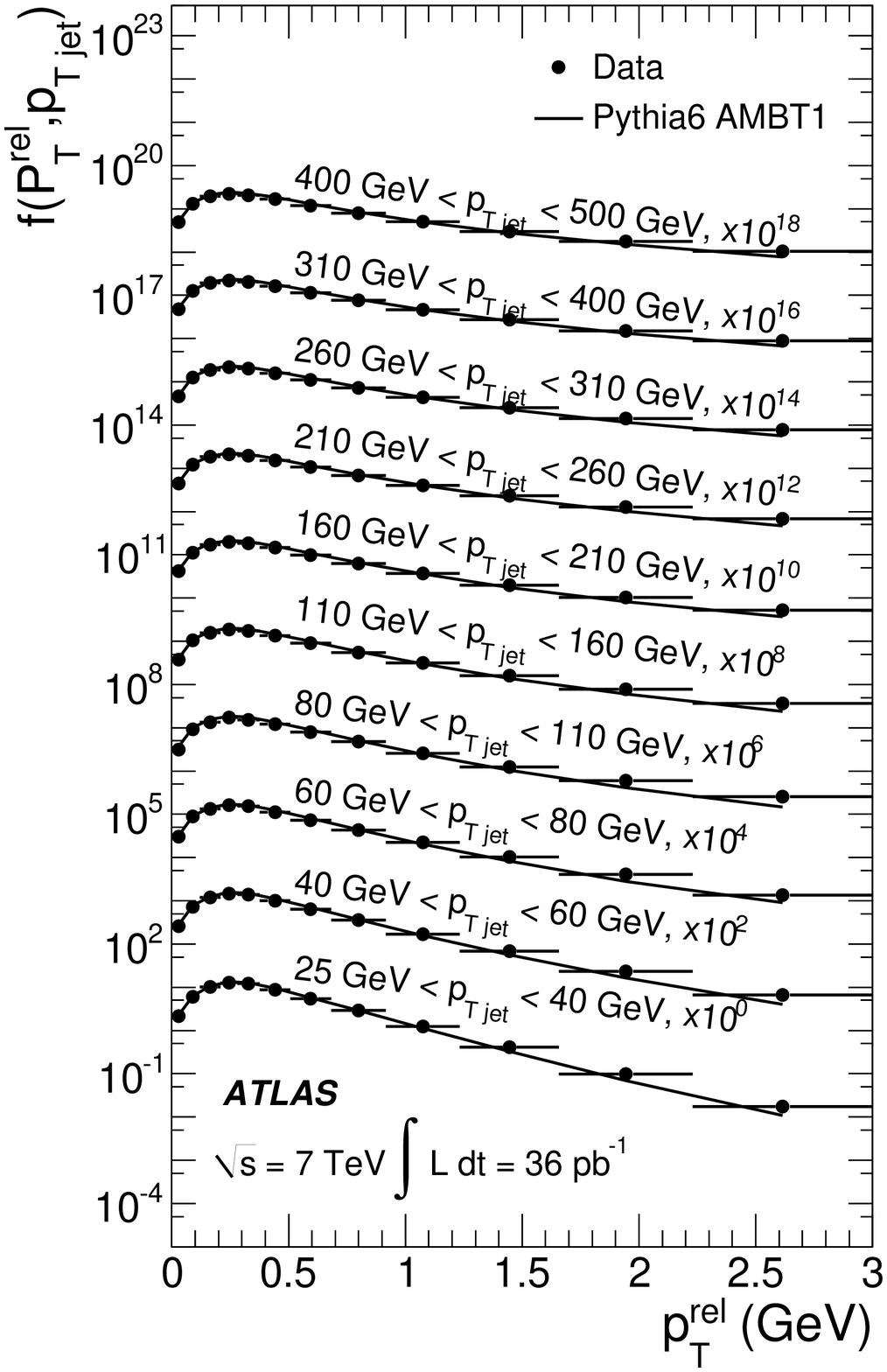}
  \end{center}
    \caption{Distributions of $f(\ptrel)$. The circles show unfolded data.  The 
lines are the predictions from AMBT1 \Pythia. 
}
  \label{Fig:ptRelProjections}
\end{figure}

\begin{figure*}
    \begin{center}
   \includegraphics[width=0.48\textwidth]{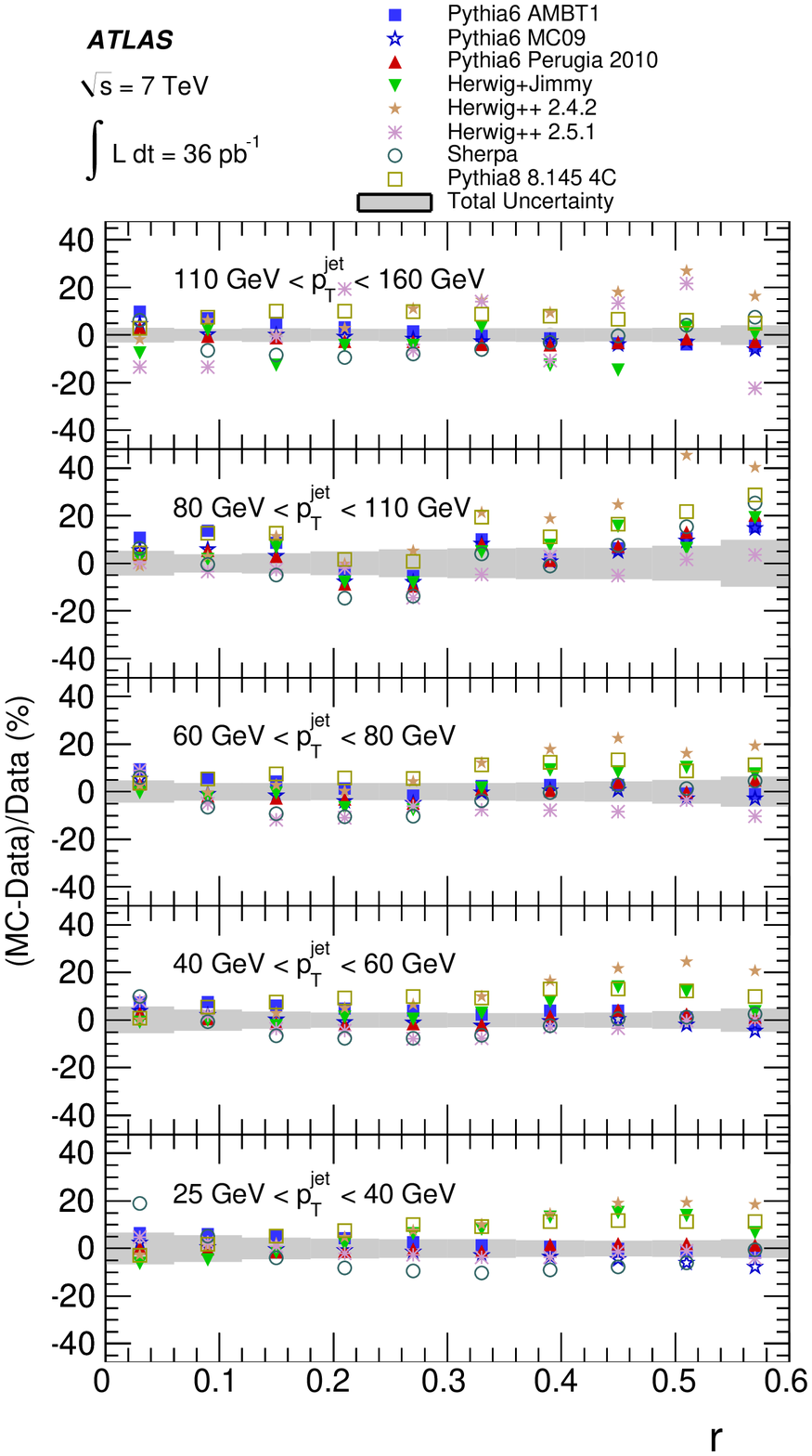}
   \includegraphics[width=0.48\textwidth]{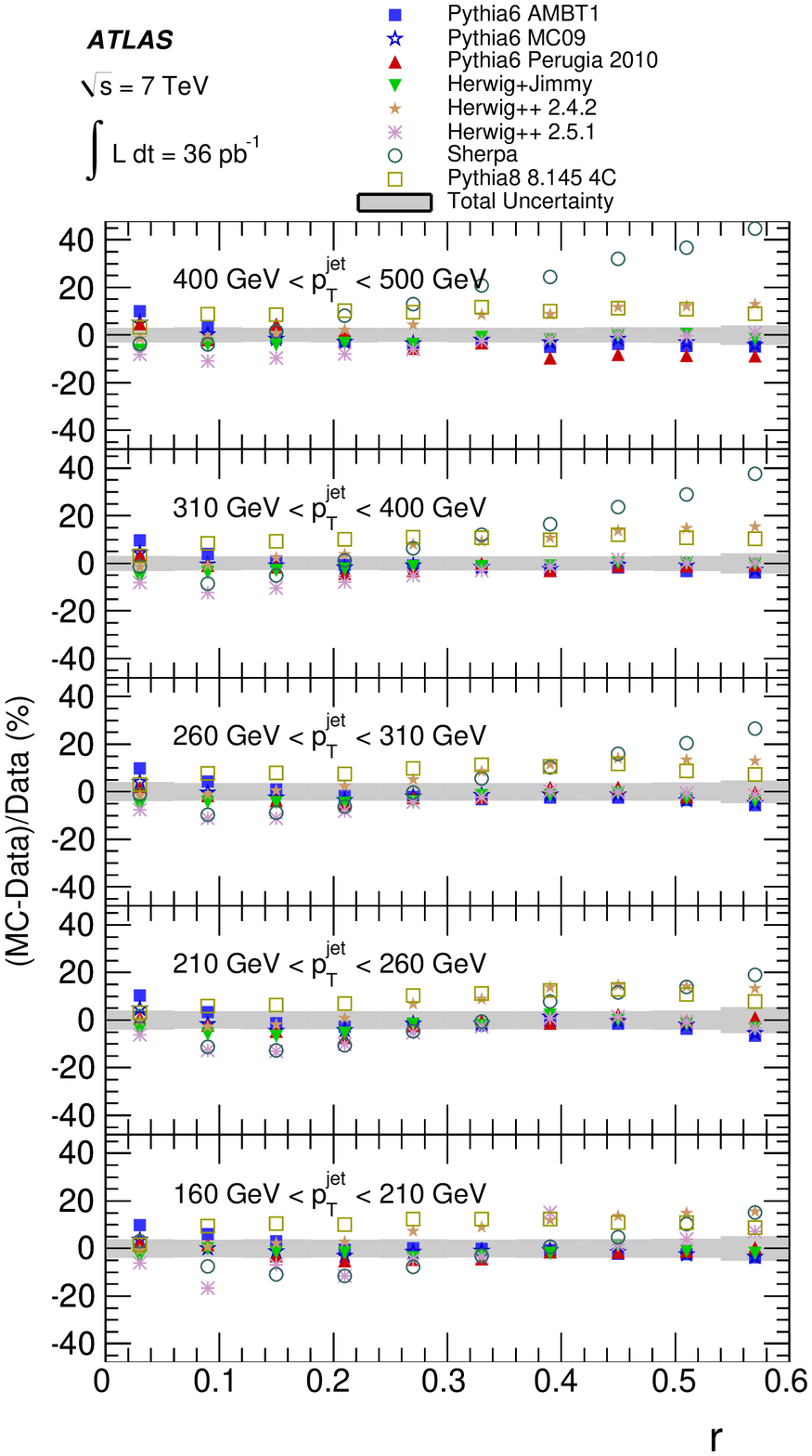}
    \end{center}
\caption{The ratio of \rhoch\ predicted by various Monte Carlo generators
to that measured.
The gray band indicates the combined statistical and systematic
uncertainties.}
\label{Fig:rhoTruthOver0}
\end{figure*}

\begin{figure*}
    \begin{center}
   \includegraphics[width=0.48\textwidth]{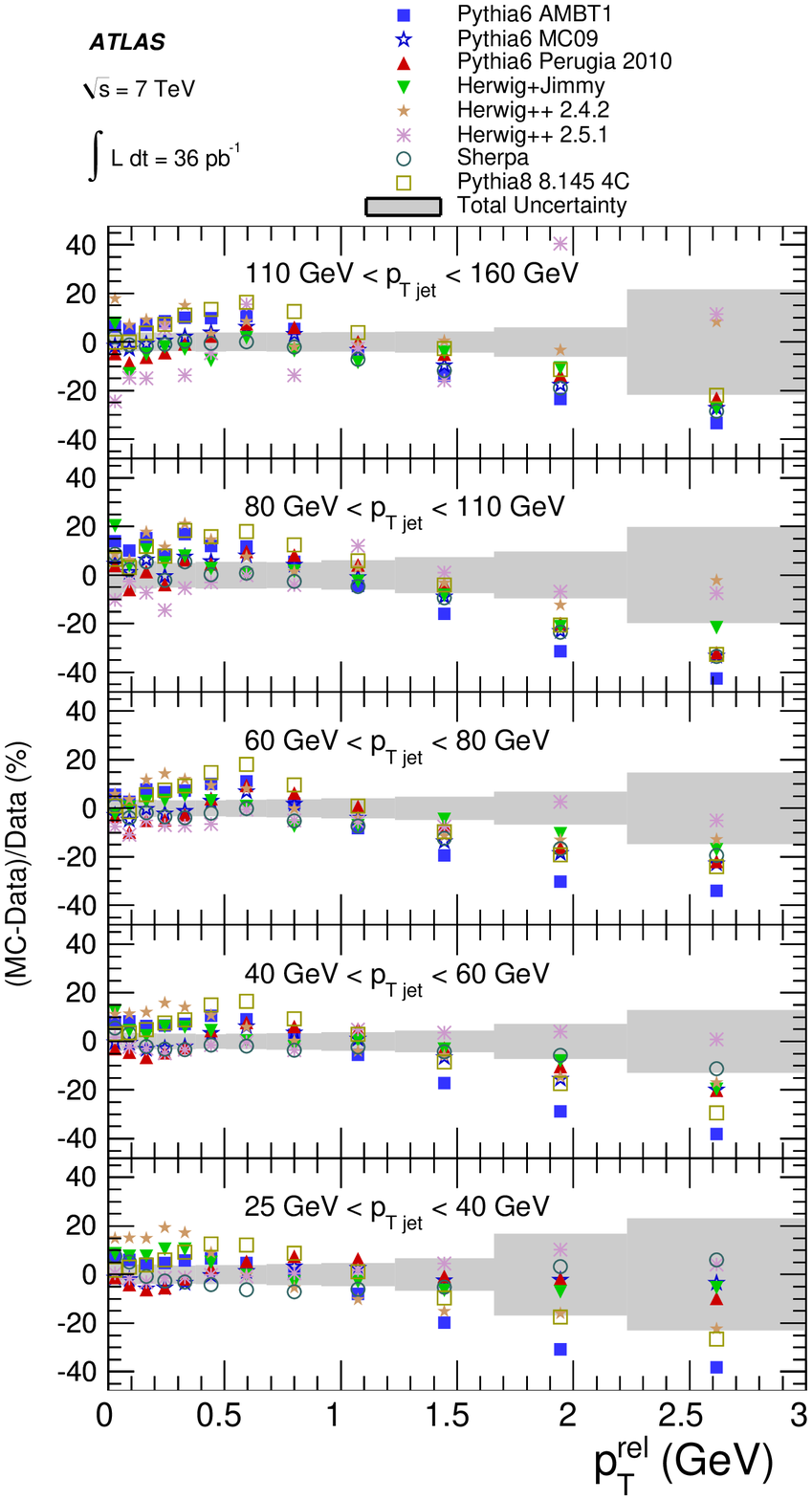}
   \includegraphics[width=0.48\textwidth]{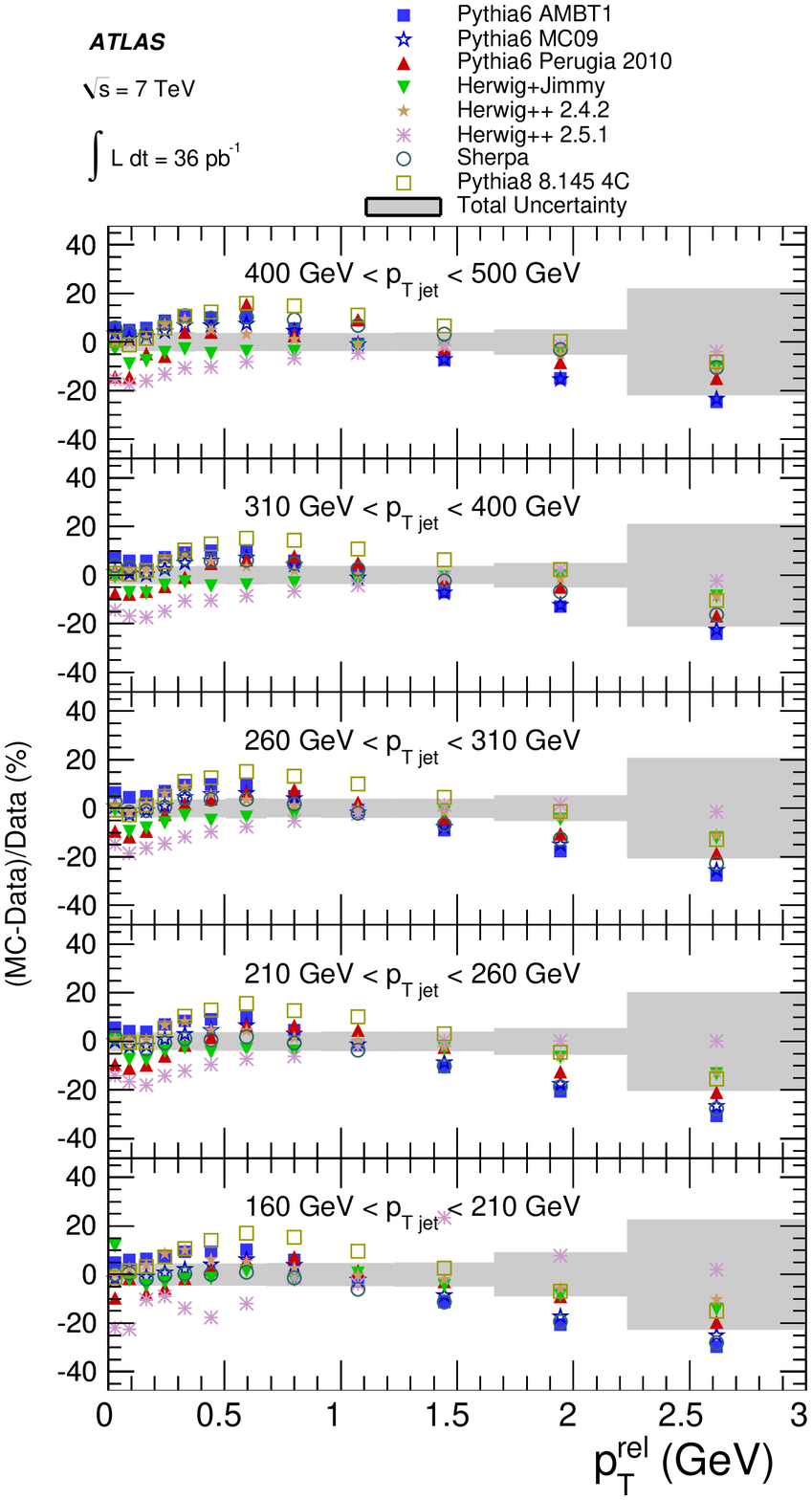}
    \end{center}
\caption{The ratio of $f(\ptrel )$ predicted by various Monte Carlo generators
to that measured.
The gray band indicates the combined statistical and systematic
uncertainties.}
\label{Fig:ptrelTruthOver0}
\end{figure*}

The transverse profile of the jets is described by the 
\rhoch\ and $f(\ptrel)$ distributions.
Figure~\ref{Fig:UnfoldingDataProjectionsrho} 
shows the distribution of \rhoch\ in two bins of \ptJet. 
The sharp decrease in population in the last bin 
is a feature of the jet algorithm, which tends to 
incorporate particles close to the radius parameter into the jet.
The effect is also seen 
in \cite{Aad:2011gn} (Figure 6)  where distributions for two radius 
parameters are shown.
Figure~\ref{Fig:UnfoldingDataProjectionsptrel} shows the distribution
of $f(\ptrel)$
in the same two \ptJet\ bins.
Figures~\ref{Fig:rhoProjections} and~\ref{Fig:ptRelProjections}
show distributions 
of  $\rho_{ch}(r)$  and $f(\ptrel)$, respectively,
in all \ptJet\ bins together with the predictions of the
AMBT1  Monte Carlo. 
Comparisons of \rhoch\ for all data and Monte Carlo
are shown in Figure~\ref{Fig:rhoTruthOver0}. 
Sherpa, \Herwigpp\ 2.4.2 and \Pythia8 disagree
significantly with the data over the full range of the measurement.
\Pythia8 \ is consistent with the data only over a very restricted range of
\ptJet\ around $80\ \GeV$. \Herwigpp\ 2.5.1 
shows good agreement except at small $r$ and for
$\ptJet> 200\ \GeV$.
Herwig+Jimmy is consistent with the data only for  $\ptJet> 160\ \GeV$. 
All the 
\Pythia~6 tunings except AMBT1 agree; AMBT1 shows disagreement for  
$\ptJet > 200\ \GeV$.
Comparison of $f(\ptrel)$ for all data and Monte Carlos
are shown in Figure~\ref{Fig:ptrelTruthOver0}.
None of the generators agree with the data within the systematic uncertainties.

\begin{figure}
  \begin{center}
   \includegraphics[width=0.48\textwidth]{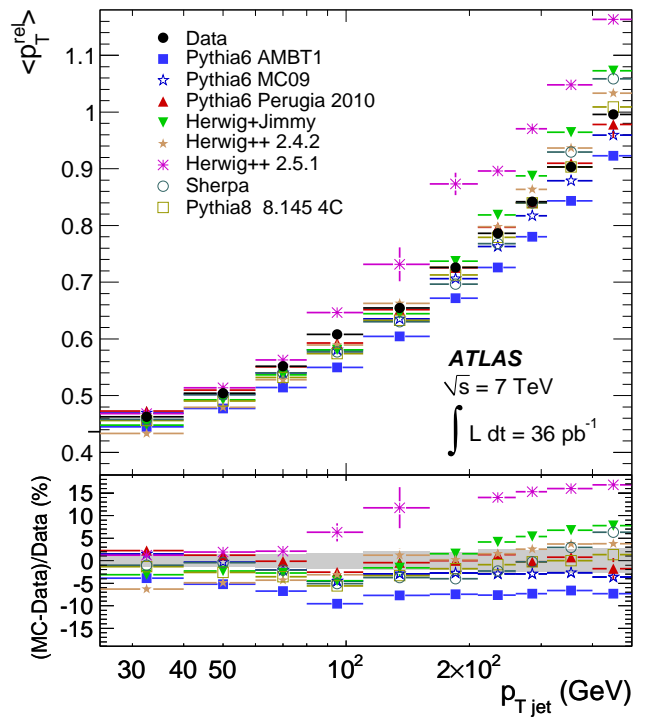}
  \end{center}
    \caption{Comparison of the measured value of the average value of 
\ptrel\  as a function of  $\ptJet$ with various Monte
Carlo expectations.}
  \label{Fig:ptrelaverage}
\end{figure}

The mean value of \ptrel\  as a function of $\ptJet$ is shown in Figure~\ref{Fig:ptrelaverage}.
\Herwigpp\ 2.5.1 has much too large a value of  $\langle \ptrel \rangle$ for  $\ptJet> 100\ \GeV$  and 2.4.2 has too small
a value for   $\ptJet< 80\ \GeV$. AMBT1 has too small a value at all  
\ptJet. Herwig+Jimmy has too large a value for
$\ptJet> 200\ \GeV$. Agreement of the remaining Monte Carlos is quite good.

\clearpage

\clearpage
\section{Conclusion}
\label{sec:conclusions}
A measurement of the jet fragmentation properties for charged particles in proton-proton collisions at a
center-of-mass energy of 7~$\TeV$ is presented. 
The dataset recorded  with the ATLAS detector at the LHC  in 2010 with an integrated luminosity of $36\,$pb$^{-1}$ is used. 
Systematic uncertainties for the
fragmentation function which describes how the jet momentum is distributed amongst its constituents 
  vary between approximately 4\%\ and 40\%\
depending on $z$ and \ptJet.  The uncertainties increase strongly  with  $z$ and are largest at small
 \ptJet.
The measurements of the  distributions
 $\rho_{ch}(r,\ptJet)$ and $f(\ptrel,\ptJet)$ which describe the shape of jets transverse to the jet direction
have uncertainties that fall as  \ptJet\  increases, increase at large values 
of
\ptrel\ and are almost independent of $r$. They are less than 5\%\ except in the lowest  \ptJet\ range and for
$\ptrel > 1 \;\GeV$.

The measurements are sensitive to several properties of QCD as implemented in and modeled by
Monte Carlo event generators. The additional QCD radiation present as  \ptJet\ increases is modeled by 
perturbative QCD and results in a growth of the particle multiplicity. This growth is very well modeled
by all the Monte Carlo generators used here.
Two other effects that
cannot be described by perturbative QCD impact the measured distributions. 
The hadronization of partons produced in a QCD radiative shower into the observed hadrons must be modeled in
the Monte Carlo generators and is described by a large number of parameters which are tuned to agree with data.
Particles produced from remnants of the initial protons (underlying event) can be incorporated into jets whose constituents mainly come from the hard 
scattering,
so the measured jet properties can be sensitive to this modeling.

The measured fragmentation functions agree well 
with the AMBT1 \Pythia\ and \Perugiatwenty\
Monte Carlo predictions within statistical and systematic uncertainties.
Other tunes and generators show less good agreement indicating that 
the non-perturbative physics is not adequately modeled in these cases.
Measurements of the transverse distributions 
$f(\ptrel,\ptJet)$  and $\rho_{ch}(r,\ptJet)$ are also presented.   
For the \ptrel\ distribution, none of 
the generators agree with data  within systematic uncertainties
over the full kinematic range.  For the $\rho_{ch}(r,\ptJet)$
distribution, Herwig+Jimmy, \Pythia~MC09 and \Perugiatwenty\ are 
in reasonable agreement  with the data. 

In summary, none of the Monte Carlo
generators studied provide a good description of all the data.
The measurements presented here provide valuable inputs to constrain future
improvements in Monte Carlo modeling of fragmentation.
The full results  are available in the
 HEPDATA database\cite{Buckley:2010jn}, and a Rivet\cite{Buckley:2010ar} module for the analysis is also  available.
\section{Acknowledgments}
We honour the memory of our young colleague Christoph Ruwiedel, who was
closely involved in the work described here and died shortly before its
completion.
We thank CERN for the very successful operation of the LHC, as well as the
support staff from our institutions without whom ATLAS could not be
operated efficiently.

We acknowledge the support of ANPCyT, Argentina; YerPhI, Armenia; ARC,
Australia; BMWF, Austria; ANAS, Azerbaijan; SSTC, Belarus; CNPq and FAPESP,
Brazil; NSERC, NRC and CFI, Canada; CERN; CONICYT, Chile; CAS, MOST and
NSFC, China; COLCIENCIAS, Colombia; MSMT CR, MPO CR and VSC CR, Czech
Republic; DNRF, DNSRC and Lundbeck Foundation, Denmark; ARTEMIS, European
Union; IN2P3-CNRS, CEA-DSM/IRFU, France; GNAS, Georgia; BMBF, DFG, HGF, MPG
and AvH Foundation, Germany; GSRT, Greece; ISF, MINERVA, GIF, DIP and
Benoziyo Center, Israel; INFN, Italy; MEXT and JSPS, Japan; CNRST, Morocco;
FOM and NWO, Netherlands; RCN, Norway; MNiSW, Poland; GRICES and FCT,
Portugal; MERYS (MECTS), Romania; MES of Russia and ROSATOM, Russian
Federation; JINR; MSTD, Serbia; MSSR, Slovakia; ARRS and MVZT, Slovenia;
DST/NRF, South Africa; MICINN, Spain; SRC and Wallenberg Foundation,
Sweden; SER, SNSF and Cantons of Bern and Geneva, Switzerland; NSC, Taiwan;
TAEK, Turkey; STFC, the Royal Society and Leverhulme Trust, United Kingdom;
DOE and NSF, United States of America.

The crucial computing support from all WLCG partners is acknowledged
gratefully, in particular from CERN and the ATLAS Tier-1 facilities at
TRIUMF (Canada), NDGF (Denmark, Norway, Sweden), CC-IN2P3 (France),
KIT/GridKA (Germany), INFN-CNAF (Italy), NL-T1 (Netherlands), PIC (Spain),
ASGC (Taiwan), RAL (UK) and BNL (USA) and in the Tier-2 facilities
worldwide.

\bibliographystyle{atlasstylem}
\bibliography{fragmentationPaper}

\begin{mcbibliography}{10}

\bibitem{Gribov:1972rt}
V.N. Gribov and L.N. Lipatov,
\newblock  Sov. J. Nucl. Phys. {\bf 15}  (1972) 675--684\relax
\relax
\bibitem{Altarelli:1977zs}
G. Altarelli and G.Parisi,
\newblock  Nucl. Phys. {\bf B126}  (1977) 298\relax
\relax
\bibitem{Dokshitzer:1977sg}
Y.L Dokshitzer,
\newblock  Sov. Phys. JETP {\bf 46}  (1977) 641--653\relax
\relax
\bibitem{Georgi:1977mg}
H. Georgi and H.D. Politzer,
\newblock  Nucl. Phys. {\bf B136}  (1978) 445\relax
\relax
\bibitem{Owens:1978qz}
J.F. Owens,
\newblock  Phys. Lett. {\bf B76}  (1978) 85\relax
\relax
\bibitem{Uematsu:1978yw}
T. Uematsu,
\newblock  Phys. Lett. {\bf B79}  (1978) 97\relax
\relax
\bibitem{Sjostrand:2006za}
T. Sj{\"{o}}strand, S. Mrenna and P.Z. Skands,
\newblock  JHEP {\bf 05}  (2006) 026\relax
\relax
\bibitem{Nakamura:2010zzi}
K. Nakamura et al.(Particle Data Group),
\newblock  J. Phys. {\bf G37}  (2010) 075021\relax
\relax
\bibitem{Mueller:1981ex}
A.H. Mueller,
\newblock  Phys. Lett. {\bf B104}  (1981) 161--164\relax
\relax
\bibitem{Corcella:2000bw}
G. Corcella et al.,
\newblock  JHEP {\bf 01}  (2001) 010\relax
\relax
\bibitem{Aad:2011gn}
The ATLAS Collaboration,
\newblock  {Properties of jets from tracks in proton-proton collisions at
  center-of-mass energy $\sqrt{s} = 7$~TeV with the ATLAS detector},
\newblock  {CERN-PH-EP-2011-110, accepted by Phys. Rev. D}, 2011\relax
\relax
\bibitem{Akers:1995wt}
The OPAL Collaboration,
\newblock  Z. Phys. {\bf C68}  (1995) 203--214\relax
\relax
\bibitem{Abreu:1996na}
The DELPHI Collaboration,
\newblock  Z. Phys. {\bf C73}  (1996) 11--60\relax
\relax
\bibitem{Buskulic:1995aw}
The ALEPH Collaboration,
\newblock  Phys. Lett. {\bf B357}  (1995) 487--499\relax
\relax
\bibitem{Mattig:1988sv}
P. M{\"{a}}ttig,
\newblock  Phys. Rept. {\bf 177}  (1989) 141\relax
\relax
\bibitem{Arnison:1983gv}
G. Arnison et al.,
\newblock  Phys. Lett. {\bf B132}  (1983) 223\relax
\relax
\bibitem{Abe:1990ii}
The CDF Collaboration,
\newblock  Phys. Rev. Lett. {\bf 65}  (1990) 968--971\relax
\relax
\bibitem{Breitweg:1997ra}
ZEUS Collaboration,
\newblock  Phys. Lett. {\bf B414}  (1997) 428--443\relax
\relax
\bibitem{Aaron:2007ds}
H1 Collaboration,
\newblock  Phys. Lett. {\bf B654}  (2007) 148--159\relax
\relax
\bibitem{ATLAS}
The ATLAS Collaboration,
\newblock  JINST {\bf 3}  (2008) S08003\relax
\relax
\bibitem{ATLAS-CONF-2011-032}
The ATLAS Collaboration,
\newblock  {Jet energy scale and its systematic uncertainty in proton-proton
  collisions at $\sqrt{s} = 7$~TeV with ATLAS 2010 data },
\newblock  ATLAS-CONF-2011-032\\
  \texttt{https://cdsweb.cern.ch/record/1337782}\relax
\relax
\bibitem{bib:jet}
The ATLAS Collaboration,
\newblock  {Eur. Phys. J. C} {\bf 71}  (2011) 1512\relax
\relax
\bibitem{bib:atlas-sim}
{The ATLAS Collaboration},
\newblock  Eur. Phys. J C {\bf 70}  (2010) 823--874\relax
\relax
\bibitem{bib:g4}
{GEANT4 Collaboration, S. Agostinelli et al.},
\newblock  Nucl. Instr. Meth. {\bf A506}  (2003) 250--303\relax
\relax
\bibitem{Sherstnev:2007nd}
A. Sherstnev and R.S. Thorne,
\newblock  Eur. Phys. J. {\bf C55}  (2008) 553--575\relax
\relax
\bibitem{Aad:2010ir}
The ATLAS Collaboration,
\newblock  New J. Phys. {\bf 13}  (2011) 053033\relax
\relax
\bibitem{Skands:2010ak}
P.Z. Skands,
\newblock  Phys. Rev. {\bf D82}  (2010) 074018\relax
\relax
\bibitem{Pumplin:2002vw}
J. Pumplin et al.,
\newblock  JHEP {\bf 07}  (2002) 012\relax
\relax
\bibitem{Butterworth:1996zw}
J.M. Butterworth, J.R. Forshaw and M.H. Seymour,
\newblock  Z. Phys. {\bf C72}  (1996) 637--646\relax
\relax
\bibitem{Bahr:2008pv}
M. Bahr et al.,
\newblock  Eur. Phys. J. {\bf C58}  (2008) 639--707\relax
\relax
\bibitem{ATL-PHYS-PUB-2010-002}
The ATLAS Collaboration,
\newblock  ATLAS Monte Carlo tunes for MC09,
\newblock  ATL-PHYS-PUB-2010-002\\
  \texttt{http://cdsweb.cern.ch/record/1247375/files/ATL-PHYS-PUB-2010-002.pdf%
}\relax
\relax
\bibitem{Gieseke:2011na}
S. Gieseke et al.,
\newblock  {Herwig++ 2.5 Release Note},
\newblock  2011\relax
\relax
\bibitem{Gleisberg:2008ta}
T. Gleisberg et al.,
\newblock  JHEP {\bf 02}  (2009) 007\relax
\relax
\bibitem{Pumplin:2005rh}
J. Pumplin, A. Belyaev, J. Huston, D. Stump and W.K. and Tung,
\newblock  JHEP {\bf 02}  (2006) 032\relax
\relax
\bibitem{Sjostrand:2007gs}
T. Sj{\"{o}}strand, S. Mrenna and P.Z. Skands,
\newblock  Comput. Phys. Commun. {\bf 178}  (2008) 852--867\relax
\relax
\bibitem{bib:antikt2}
{M. Cacciari, G.P. Salam and G. Soyez},
\newblock  JHEP {\bf 04}  (2008) 063\relax
\relax
\bibitem{bib:antikt3}
{M. Cacciari and G.P. Salam},
\newblock  Phys. Lett. B {\bf 641}  (2006) 057\relax
\relax
\bibitem{bib:topo}
{W. Lampl, et. al},
\newblock  {Calorimeter clustering algorithms: Description and performance},
\newblock  ATL-LARG-PUB-2008-002\\
  \texttt{https://cdsweb.cern.ch/record/1099735}\relax
\relax
\bibitem{D'Agostini:1994zf}
G. D'Agostini,
\newblock  Nucl. Instrum. Meth. {\bf A362}  (1995) 487--498\relax
\relax
\bibitem{RooUnfold}
T. Adye,
\newblock  {RooUnfold: unfolding framework and algorithms},
\newblock  presentations to the Oxford and RAL ATLAS Groups May 2008\\
  \texttt{http://hepunx.rl.ac.uk/{\raise.17ex\hbox{$\scriptstyle\sim$}}adye/ta%
lks/}\texttt{080520-unfold.pdf}\relax
\relax
\bibitem{ATLAS-CONF-2010-054}
{The ATLAS Collaboration},
\newblock  {Jet energy resolution and reconstruction efficiencies from in-situ
  techniques with the ATLAS Detector Using Proton-Proton Collisions at a Center
  of Mass Energy $\sqrt{s} = 7$~TeV},
\newblock  ATLAS-CONF-2010-054\relax
\relax
\bibitem{mb2paper}
The ATLAS Collaboration,
\newblock  New J. Phys. {\bf 13}  (2011) 053033\relax
\relax
\bibitem{ATLAS-CONF-2011-003}
{The ATLAS Collaboration},
\newblock  {Muon Momentum Resolution in First Pass Reconstruction of pp
  Collision Data Recorded by ATLAS in 2010},
\newblock  ATLAS-CONF-2011-003 \\
  \texttt{http://cdsweb.cern.ch/record/1322424}\relax
\relax
\bibitem{ATLAS-CONF-2011-028}
{The ATLAS Collaboration},
\newblock  {ATLAS Calorimeter Response to Single Isolated Hadrons and
  Estimation of the Calorimeter Jet Scale Uncertainty},
\newblock  ATLAS-CONF-2011-028\\
  \texttt{http://cdsweb.cern.ch/record/1337075}\relax
\relax
\bibitem{erfon:bootstrap}
B. Efron,
\newblock  Annals of Statistics {\bf 7}  (1979) 1\relax
\relax
\bibitem{Buckley:2010jn}
A. Buckley and M. Whalley,
\newblock  PoS {\bf ACAT2010}  (2010) 067\relax
\relax
\bibitem{Buckley:2010ar}
A. Buckley et al.,
\newblock  arXiv: {\bf 1003.0694}  (2010)\relax
\relax
\end{mcbibliography}

\clearpage
\onecolumn
\begin{flushleft}
{\Large The ATLAS Collaboration}

\bigskip

G.~Aad$^{\rm 48}$,
B.~Abbott$^{\rm 111}$,
J.~Abdallah$^{\rm 11}$,
A.A.~Abdelalim$^{\rm 49}$,
A.~Abdesselam$^{\rm 118}$,
O.~Abdinov$^{\rm 10}$,
B.~Abi$^{\rm 112}$,
M.~Abolins$^{\rm 88}$,
H.~Abramowicz$^{\rm 153}$,
H.~Abreu$^{\rm 115}$
% ...
\bigskip

% ...
$^{10}$ Institute of Physics, Azerbaijan Academy of Sciences, Baku, Azerbaijan\\
$^{11}$ Institut de F\'isica d'Altes Energies and Departament de F\'isica de la Universitat Aut\`onoma  de Barcelona and ICREA, Barcelona, Spain\\
% ...
$^{48}$ Fakult\"{a}t f\"{u}r Mathematik und Physik, Albert-Ludwigs-Universit\"{a}t, Freiburg i.Br., Germany\\
$^{49}$ Section de Physique, Universit\'e de Gen\`eve, Geneva, Switzerland\\
% ...
$^{88}$ Department of Physics and Astronomy, Michigan State University, East Lansing MI, United States of America\\
% ...
$^{111}$ Homer L. Dodge Department of Physics and Astronomy, University of Oklahoma, Norman OK, United States of America\\
$^{112}$ Department of Physics, Oklahoma State University, Stillwater OK, United States of America\\
% ...
$^{115}$ LAL, Univ. Paris-Sud and CNRS/IN2P3, Orsay, France\\
% ...
$^{118}$ Department of Physics, Oxford University, Oxford, United Kingdom\\
% ...
$^{153}$ Raymond and Beverly Sackler School of Physics and Astronomy, Tel Aviv University, Tel Aviv, Israel\\
% ...
\end{flushleft}


\end{document}